\documentclass{WileyMSP-template}
\usepackage{soul}
\usepackage{xcolor}
\sethlcolor{yellow}
\usepackage[utf8]{inputenc}
\usepackage{textgreek}
\usepackage{tabularx}

\begin{document}

\pagestyle{fancy}
\rhead{\includegraphics[width=2.5cm]{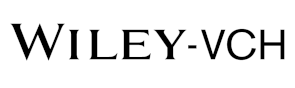}}

\title{Engineering 2D Van der Waals Electrode via MBE-Grown Weyl Semimetal 1T$'$-WTe$_{2}$ for Enhanced Photodetection in InSe}

\maketitle


\author{Biswajit Khan,}
\author{Santanu Kandar,}
\author{Taslim Khan,}
\author{Kritika Bhattacharya,}
\author{Nahid Chowdhury,}
\author{Suprovat Ghosh,}
\author{Pawan Kumar,}
\author{Rajendra Singh* and}
\author{Samaresh Das*}

\begin{affiliations}
Biswajit Khan, Kritika Bhattacharya, Suprovat Ghosh and Pawan Kumar \\
Centre for Applied Research in Electronics,\\ 
Indian Institute of Technology Delhi,\\
New Delhi 110016, India. \\
E-mail: biswajit.khan@care.iitd.ac.in\\
\vspace{4mm}
Santanu Kandar and Taslim Khan \\
Department of Physics,\\ 
Indian Institute of Technology Delhi,\\
New Delhi 110016, India. \\
E-mail: santanu.kandar@phy.iitd.ac.in\\

\vspace{4mm}
Nahid Chowdhury\\
The School of Interdisciplinary Research,\\ 
Indian Institute of Technology Delhi,\\
New Delhi 110016, India. \\

\vspace{4mm}
Rajendra Singh \\
Department of Physics,\\ 
Indian Institute of Technology Delhi,\\
New Delhi 110016, India. \\
E-mail: rsingh@physics.iitd.ac.in\\

\vspace{4mm}
Samaresh Das \\
Centre for Applied Research in Electronics,\\ 
Indian Institute of Technology Delhi,\\
New Delhi 110016, India. \\
E-mail: samaresh.das@care.iitd.ac.in\\
\end{affiliations}

\begin{abstract}
\begin{justify}
\textbf{Achieving low contact resistance in advanced quantum electronic devices remains a critical challenge. With the growing demand for faster and energy-efficient devices, 2D contact engineering offers a promising solution. Beyond graphene, 1T$'$-WTe$_2$ has attracted attention for its excellent electrical transport, quantum phenomena, and Weyl semimetallic properties. Here, we demonstrate the direct wafer-scale growth of 1T$'$-WTe$_2$ via molecular beam epitaxy (MBE) and its use as a 2D contact for layered materials such as InSe. The 1T$'$-WTe$_2$/InSe interface exhibits a barrier height nearly half that of conventional metal contacts, and its contact resistance is reduced by a factor of 21, effectively suppressing Fermi-level pinning and enabling efficient electron injection. InSe/1T$'$-WTe$_2$ photodetectors show broad photoresponsivity (0.14--217.58~A/W) under NIR to DUV illumination with fast rise/fall times of 42/126~ms, compared to lower responsivity (8.65~$\times$~10$^{-4}$--3.64~A/W) and slower response (150/144~ms) for InSe/Ti--Au devices. The 1T$'$-WTe$_2$/InSe devices thus exhibit $\sim$60$\times$ higher responsivity and $\sim$4$\times$ faster response than conventional metal contacts. These results establish MBE-grown 1T$'$-WTe$_2$ as an effective 2D electrode, enhancing photodetection performance while simplifying device architecture, making it a strong candidate for next-generation nanoelectronic and optoelectronic devices.}
\end{justify}
\end{abstract}


\section*{\textbf{Graphical Abstract}}
\begin{figure}[H]
        \renewcommand{\figurename}{\textbf{Figure }}
           \renewcommand{\thefigure}{\textbf{2}}
            \centering
	\includegraphics[scale=.8]{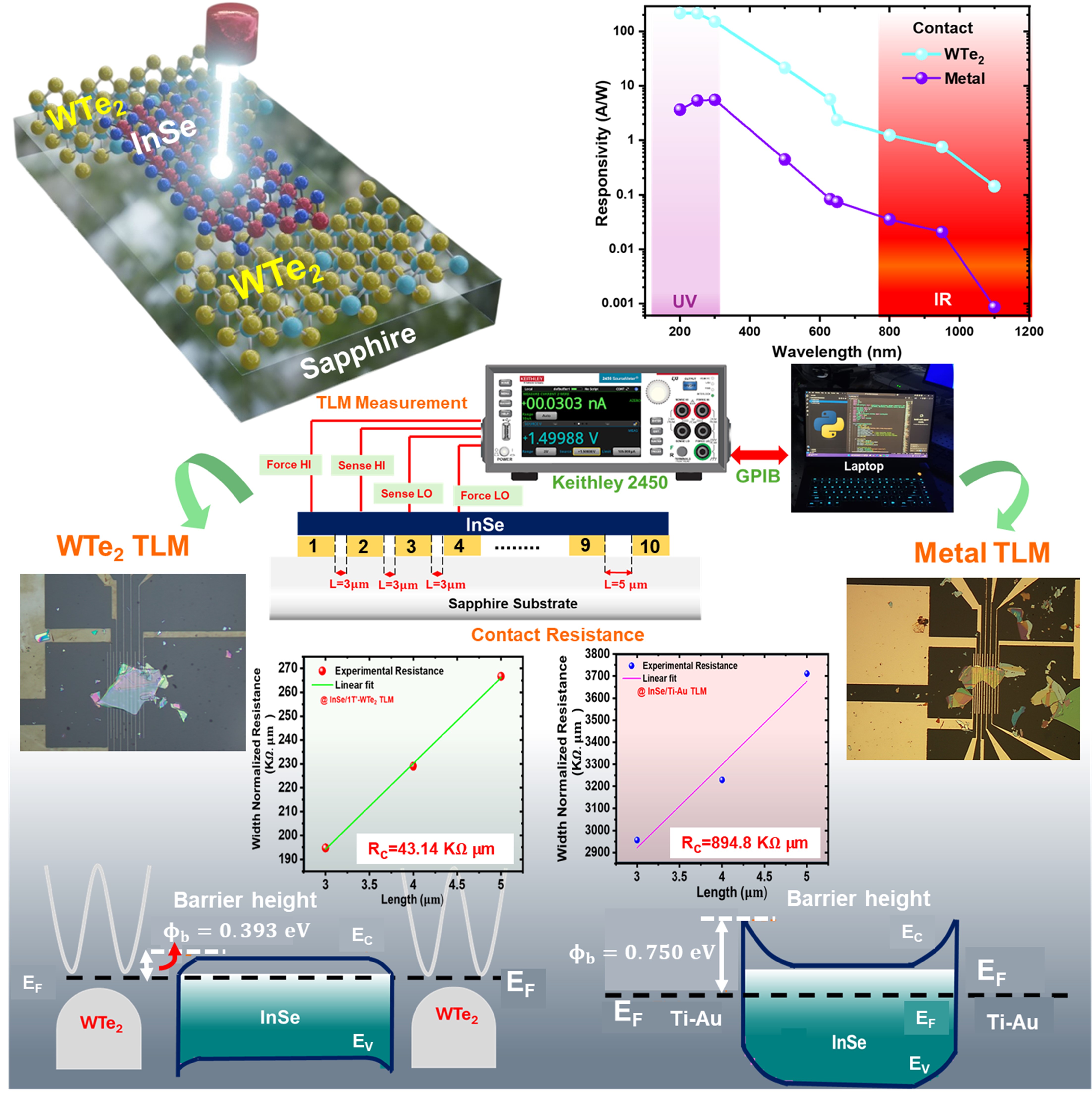}
		
\end{figure}

\section*{\textbf{TOC Summary:}}
\begin{justify}
The 1T$'$-WTe$_2$ contact, a van der Waals Weyl semimetal, significantly reduces the barrier height on InSe, which is about half that of conventional metal contacts. The contact resistance of 1T$'$-WTe$_2$ is twenty-one times lower than that of conventional metal contacts. As a result, the 1T$'$-WTe$_2$/InSe device exhibits a responsivity that is sixty times higher and four times faster than conventional metal-based devices. This study demonstrates that 1T$'$-WTe$_2$ is an excellent contact material for 2D quantum electronics due to its exceptional ability to suppress Fermi-level pinning and enable efficient electron injection by reducing both the barrier height and the contact resistance.
\end{justify}

\section*{Table of Contents}

\begin{tabularx}{\textwidth}{@{}Xr@{}}
Introduction \dotfill & 3 \\[2pt]

Materials Growth \dotfill & 5 \\
\hspace{1em} 2.1 \quad MBE growth of 1T$'$-WTe$_2$ and characterization \dotfill & 5 \\[2pt]

Heterostructure Characterisation and Device Fabrication \dotfill & 6 \\
\hspace{1em} 3.1 \quad Atomic Force Microscopy (AFM) and X-ray Diffraction (XRD) \dotfill & 6 \\
\hspace{1em} 3.2 \quad X-ray Photoelectron Spectroscopy (XPS) \dotfill & 7 \\
\hspace{1em} 3.3 \quad Device fabrication of 1T$'$-WTe$_2$/InSe/1T$'$-WTe$_2$ and Ti-Au/InSe/Ti-Au \dotfill & 9 \\
\hspace{1em} 3.4 \quad Barrier height estimation and Fermi level pinning \dotfill & 9 \\
\hspace{1em} 3.5 \quad Experimental extraction of contact resistance \dotfill & 12 \\[2pt]

Experimental Photoresponse Measurements \dotfill & 14 \\
\hspace{1em} 4.1 \quad Optoelectronic Response Analysis \dotfill & 14 \\
\hspace{1em} 4.2 \quad Response Time Analysis \dotfill & 15 \\[2pt]

Conclusion \dotfill & 18 \\[2pt]

Experimental Section \dotfill & 21 \\
\hspace{1em} 6.1 \quad Materials Growth and Processing \dotfill & 21 \\
\hspace{1em} 6.2 \quad Etching and Device Fabrication Details \dotfill & 21 \\
\hspace{1em} 6.3 \quad Characterization Techniques \dotfill & 21 \\
\hspace{2em} 6.3.1 \quad Surface Topography and KPFM \dotfill & 21 \\
\hspace{2em} 6.3.2 \quad Phase and Composition \dotfill & 21 \\
\hspace{1em} 6.4 \quad Optical and Electrical Properties \dotfill & 21 \\[2pt]

Supporting Information \dotfill & 22 \\
Acknowledgments \dotfill & 22 \\[2pt]

References \dotfill & 23 \\
\end{tabularx}

\section{Introduction}
\begin{justify}
Two-dimensional (2D) van der Waals electrodes are gaining widespread attention as the miniaturization of modern transistors amplifies the issue of contact resistance at the metal-semiconductor interface. As transistors become smaller, this interface becomes a critical bottleneck, restricting further downsizing and impacting overall device efficiency. This problem is even more significant for ultra-thin 2D semiconductors intended for next-generation nanoelectronic and optoelectronic devices. Poor contact resistance ultimately limits device performance. A substantial Schottky barrier is one of the main reasons for poor contact in metal\slash 2D semiconductor interfaces. In such scenarios, van der Waals electrodes, with their atomically thin, smooth surfaces and weak interlayer van der Waals forces, offer a refined solution by enabling low-resistance, seamless contacts, overcoming many of the challenges inherent in conventional metal-semiconductor junctions. Layered materials such as dirac and weyl semimetal (e.g. graphene, 1T$^{'}$-WTe$_{2}$) worked as  an excellent 2D van der waals electrode \cite{liu2021transferred,cusati2017electrical,mootheri2020graphene}. Such electrode not only solve the contact resitance problems but also they enhanced the speed and photoresponsivity of optoelectronic devices \cite{chen2015high}. Because building photoelectronic devices with both high speed and high responsiveness is essential for the next generation of optoelectronics. The introduction of Dirac and Weyl semimetal contacts can greatly enhance the speed of devices due to their unique dispersion relations and high mobility \cite{xia2009ultrafast,wang2012electronics,bonaccorso2010graphene, nematollahi2019weyl}. However, the inherently weak absorption and low built-in potential in these graphene \cite{xia2009ultrafast} and weyl semimetal \cite{wang2022hybrid, ma2019nonlinear, yang2023centimeter, chan2017photocurrents} based photodetectors have significantly limited their photoresponsivity. In order to increase the responsivity, layered materials like transition-metal dichalcogenides (TMDCs)\cite{wang2012electronics,yin2012single, butler2013progress} and $\mathrm{III}-\mathrm{VI}$\cite{lei2015atomically,feng2015ultrahigh} attracted tremendous attenion because of their finite bnd gap and decent photoresponsivity. Compared to other III-VI materials, bulk InSe offers a wider spectral response because of its shorter direct bandgap. Despite having wide spectral responsivity, InSe exhibited a slow response speed\cite{zhao2020role, tamalampudi2014high}. To investigate the effect of 2D van der Waals electrodes and the enhancement of speed, different groups chose InSe, which has a wider spectral response, with semimetals, especially graphene, as the contact material. 

Luo. et. al.\cite{luo2015gate} showed that semimetal like graphene enhances the speed 40 times compared to the conventional metal contact on layered material like InSe and they got photoresponsivity as high as 60 $AW^{-1}$. Chen. et. al. \cite{chen2015high} also repoted enhancement of speed ($<\SI{1}{\milli\second}$) and photoresponsivity when they used graphene/few layer of InSe. There is a significant drop in speed and responsiveness when a conductive metal, like gold (Au), is used on top of InSe. \cite{luo2015gate,tamalampudi2014high,zhao2020role}. Even the heterostructure with InSe and conventional metal contacts, such as $WSe_{2}$\slash InSe\cite{lei2021ambipolar}, Black phosphorus (BP)\slash InSe\cite{zhao2018highly}, p-InSe\slash n-InSe \cite{patil2022self}, Se\slash InSe \cite{shang2020mixed}, InSe\slash $Rse_{2}$\cite{ma2022ultrasensitive} reported low speed and low responsivity in the range of $mAW^{-1}$. Anti-ambipolar transfer is demonstrated by van der Waals (vdW) heterostructures, like p-type MoTe$_2$/n-type InSe stacks, which have a high peak-to-valley current ratio of 4103 and a broad 460\,V window. They provide self-driven photodetection at 405\,nm illumination, with a detectivity of around $3.02 \times 10^{14}$\,Jones and a photoresponsivity of 15.4\,mA/W~\cite{sun2021anti}. Polarization-sensitive detection is made possible by a GeAs/InSe heterojunction with type-II alignment, which provides $>10^3$ rectification ratio, $\sim$0.1\, pA dark current, 357\, mA/W responsivity, $10^3$ photo-switching ratio, $2 \times 10^{11}$\,Jones detectivity, 8\% PCE, and an anisotropic photocurrent ratio of $\sim$18~\cite{xiong2021high}. $>10^3$ rectification ratio, broadband response at 405-1100\,nm, 215\,mA/W responsivity, 42.2\% EQE, $1.05 \times 10^{10}$\,Jones detectivity, 8.6/4.2\, ms response time, and 4.6 anisotropic ratio for thin SnS are all demonstrated by a SnS/InSe device \cite{gao2022low}. Self-driven broadband detection (400-1064\,nm) is achieved by Bi $_2$O$_2$ Se/InSe homojunctions, with 791\,mA/W responsivity, $1.08 \times 10^{12}$\,Jones detectivity, $I_{\text{light}}/I_{\text{dark}} > 10^5$, and 5.8/15\,ms response\cite{zhang2023type}. Through a photo-gating effect, CSZCS QDs/InSe 2D-0D hybrids exhibit $1.69 \times 10^{12}$\,Jones detectivity and 300$\times$ greater responsivity (30.16\,A/W)~\cite{dan2021improved}. Ultralow $3 \times 10^{-14}$\,A dark current, $10^8$ rectification, $10^5$ on/off ratio, $1.77 \times 10^{11}$\,Jones detectivity, and 320\,$\mu$s responsiveness are all displayed by InSe/Te heterostructures\cite{liu2022selectively}. Through carrier recirculation, InSe/GaSe photodetectors can detect down to 1\,$\mu$W/cm$^2$, give $10^7$ gain, 1037\,A/W responsivity, and $8.6 \times 10^{13}$\,Jones detectivity~\cite{shang2022carrier}.
 On the other hand, 1T$'$-WTe\textsubscript{2} has been employed to engineer asymmetric Schottky junctions that enable reconfigurable photoresponse in low-dimensional heterostructures. Semimetallic 1T$'$-WTe\textsubscript{2} enables reconfigurable phototransistors via asymmetric Schottky contacts, as demonstrated in heterostructures such as 1T$'$ WTe\textsubscript{2}/WS\textsubscript{2}~\cite{ma2024vertical} and MoTe\textsubscript{2}/WTe\textsubscript{2}~\cite{xie2021gate}. Jingyi Ma \textit{et al.}~\cite{ma2024vertical} reported gate-tunable Schottky-type reconfigurable phototransistors (RPTs) with rectification ratios ranging from 10\textsuperscript{--2} to 10\textsuperscript{5}, and bidirectional photoresponse from --1325 to 430~mA/W at 635~nm. Yuan Xie \textit{et al.}~\cite{xie2021gate} demonstrated reversible rectification behavior(10\textsuperscript{2}--10\textsuperscript{5}) and switchable self-excited photocurrent in MoTe\textsubscript{2}/WTe\textsubscript{2} devices, achieving photoresponsivity up to 220~mA/W. Moreover, Jina Wang \textit{et al.}~\cite{wang2022weyl} achieved broadband (400--1100~nm) and fast photoresponse in a WTe\textsubscript{2}/GaAs Schottky diode with high responsivity (298~mA/W) and detectivity (1.7~$\times$~10\textsuperscript{12}~Jones). Luji Li \textit{et al.}~\cite{li2021tunable} further demonstrated that WTe\textsubscript{2}/MoS\textsubscript{2} heterojunctions offer stable and tunable photoresponse from the visible to near-infrared range, with improved speed and sensitivity even under low-light conditions. 
 
 As discussed in the preceding section, numerous studies have reported WTe\textsubscript{2}-based heterostructure photodetectors and phototransistors, where WTe\textsubscript{2} is typically employed in combination with other two-dimensional (2D) materials. However, utilizing 1T$^\prime$-WTe\textsubscript{2} as an electrode material, particularly in contact with semiconducting 2D layers such as InSe, presents a broader and more versatile approach beyond its conventional use in heterostructures. This functional integration enables novel device architectures and expands the potential applications of 1T$^\prime$-WTe\textsubscript{2} in next-generation optoelectronics. The growth of 1T$^\prime$-WTe\textsubscript{2} has been extensively studied, with chemical vapor deposition (CVD) being the most commonly adopted synthesis technique. For instance, Carl H. Naylor et al. demonstrated the CVD growth of monolayer 1T$^\prime$-WTe\textsubscript{2}. However, the resulting films consisted of small, non-uniform flakes, limiting their viability for scalable device applications\cite{naylor2017large}. Similarly, Yu Zhou et al. reported thickness-tunable CVD-grown 1T$^\prime$-WTe\textsubscript{2} films that were polycrystalline, reflecting challenges in achieving uniform crystallinity and large-area homogeneity \cite{zhou2017direct}. In Recent efforts, lin. et al.,\cite{lin2024dramatically} have demonstrated the feasibility of large-area CVD synthesis of 1T$^\prime$-WTe\textsubscript{2}, though issues related to film quality and reproducibility remain (citation). In contrast, molecular beam epitaxy (MBE) provides distinct advantages, including atomic-level control over thickness, superior film uniformity, and high crystalline quality over wafer-scale areas. Although only a few reports exist on the MBE growth of 1T$^\prime$-WTe\textsubscript{2}\cite{li2020molecular,walsh2017w}, this method offers enhanced reproducibility, robustness, and integration compatibility. MBE-grown 1T$^\prime$-WTe\textsubscript{2} enables reliable device fabrication with minimized variability, supporting seamless incorporation into wafer-level processes.

Here, we propose a Weyl semimetal-like 1T$^{'}$-WTe$_{2}$ as a van der Waals 2D electrode for an InSe-based wideband photodetector. Remarkably, even without utilizing a complex heterostructure, we achieve high responsivity and fast operation through the simple integration of a 1T$^{'}$-WTe$_{2}$-based 2D electrode on the layered InSe material. The introduction of 1T$^{'}$-WTe$_{2}$ not only enhances speed and responsivity but also significantly suppresses the dark current in comparison to graphene electrodes, making this approach a promising advancement for photodetection technologies.
Using a variety of methods, it was confirmed that the as-grown 2D 1T$^{'}$-WTe${2}$ had good continuity and great crystallinity. X-ray photoelectron spectroscopy (XPS) was used to confirm the chemical composition of pure 1T$^{'}$-WTe$_{2}$. X-ray diffraction and Raman confirm that the phase of molecular beam epitaxy (MBE) grown WTe${2}$ is actually 1T$^{'}$-WTe$_{2}$. A layered InSe/1T$^{'}$-WTe${2}$ structure was used to produce, test, and compare photodetectors with an InSe/Ti-Au arrangement. In addition, the devices demonstrated a high responsivity of up to 217.58 A/W, which is 60 times larger than the InSe/Ti-Au equivalent. This broadband photoresponse covered wavelengths ranging from deep ultraviolet (DUV) to near-infrared (NIR). The InSe/1T$^{'}$-WTe${2}$ configuration showed a rise time that was 4 times faster than the InSe/Ti-Au setup in terms of response time. All of these improvements arise from the suppression of Fermi-level pinning by the 1T$^{'}$-WTe${2}$ electrode. As a result, the 1T$'$-WTe$_2$/InSe device exhibits nearly half the barrier height and a contact resistance that is twenty-one times lower compared to the Ti–Au/InSe counterpart. These findings demonstrate the InSe/1T$^{'}$-WTe$_{2}$ photodiodes' notable performance improvements, which make them attractive options for high-efficiency, quick-response photodetectors. Unlike graphene, which exhibits high dark current due to its zero bandgap, 1T$'$-WTe\textsubscript{2} possesses a small intrinsic bandgap that naturally reduces leakage. This enables low-dark-current operation while maintaining good conductivity, making it an ideal van der Waals contact material for scalable, high-performance optoelectronic and quantum devices\cite{khan2024observation}. In summary, our study highlights how the integration of low-barrier contact engineering with wafer-scale, molecular beam epitaxy (MBE)-grown 1T$'$-WTe\textsubscript{2} provides a powerful pathway toward next-generation 2D optoelectronic and quantum devices. This approach achieves high responsivity, faster photoresponse, and significantly reduced leakage, while maintaining reproducibility and scalability. These attributes position 1T$'$-WTe\textsubscript{2} as a practical and high-performance contact material suitable for future energy-efficient and quantum-integrated platforms.
\end{justify}
\section{Materials Growth :} 
\subsection{\textbf{\large MBE growth of 1T$'$ WTe$_{2}$ and characterization}}
\begin{justify}
Uniform and stoichiometric growth of 1T$^\prime$-WTe\textsubscript{2} over large-area presents challenges for various growth techniques, as it requires precise control over parameters such as the tungsten (W) and tellurium (Te) flux, growth temperature, and deposition rate. Molecular beam epitaxy (MBE) is chosen for growing WTe2 over other techniques due to its ability to precisely control these growth parameters, ensuring the growth of uniform, crystalline, and stoichiometric WTe2 films with minimal defects. Tellurium (Te) is highly volatile, making it difficult to maintain its concentration during the growth process.  Additionally, achieving a good stoichiometric 1T$^\prime$-WTe\textsubscript{2} film is challenging due to the difficulty in the bond formation between tungsten (W) and tellurium. This requires precise control over both the growth temperature and the flux of the precursor materials. To address these challenges, a two-step growth process is employed for 1T$^\prime$-WTe\textsubscript{2} film deposition. In the first step, the C-plane sapphire substrate is exposed to a constant Te flux for 10 minutes to create a Te-rich surface before the arrival of W. In the second step, a continuous, optimized flux of W and Te is provided to grow the 1T$^\prime$-WTe\textsubscript{2} film as shown in Figure 1a. The growth temperature is controlled at 450°C to ensure a uniform and crystalline 1T$^\prime$-WTe\textsubscript{2} structure. Throughout the process, in-situ reflection high-energy electron diffraction (RHEED) is used to monitor the surface evolution of the film. Initially, a polycrystalline RHEED pattern is observed in the form of rings, which convert to a streaky pattern, indicating the development of a crystalline structure as the growth progresses (figure -\ref{fig1}b). The 1T$^\prime$-WTe\textsubscript{2} film, consisting of 10 layers, is grown on the sapphire substrate with a growth rate of 0.013 nm/min. This two-step approach, combined with real-time monitoring via RHEED and phase identification through Raman spectroscopy, energy-dispersive X-ray (EDX) demonstrates an effective methodology for the growth of high-quality and nearly stoichiometric WTe2 films Figure-\ref{fig1}c shows an image of the large-scale MBE-grown 1T$^{'}$-WTe$_{2}$ film. Post-growth characterization using Raman spectroscopy confirms the formation of the stable semi-metallic 1T$^{'}$ phase of 1T$^\prime$-WTe\textsubscript{2} \cite{buchkov2021anisotropic,kim2016anomalous,walsh2017w,xiao2024photoelectric}, as shown in figure-\ref{fig1}d. The observed Raman modes in Figure-\ref{fig1}d for 1T$^\prime$-WTe$_2$ correspond to the A$_{2}^{2}$, A$_{2}^{4}$, A$_{1}^{3}$, A$_{1}^{4}$, A$_{1}^{7}$, A$_{1}^{8}$, and A$_{1}^{9}$ vibrational modes, located at 88.9, 103.7, 120.2, 134.5, 168.0, 183.0, and 217.9~cm$^{-1}$, respectively. Additionally, an unassigned vibrational mode is observed at 233.5~cm$^{-1}$.EDX mapping was performed on a sample grown using MBE to analyze elemental distributions and continuity of the grown film at different positions across the sample. MBE grown sample is marked with five distinct locations (figure-\ref{fig1}c): Position-1, Position-2, Position-3, Position-4, and Position-5, where EDX data were collected. These maps offer a comprehensive view of elemental distributions across the sample and allow for comparison of compositional uniformity from the center to the periphery, helping to confirm the spatial consistency of the MBE-grown material across the entire sample. Notably, Position-5, which lies at the center of the sample, is highlighted in figure-\ref{fig1}c, displaying the elemental distribution and uniformity at the core of the growth. The EDX mapping at the center (Position-5) is shown in figure-\ref{fig1}e to \ref{fig1}g. 
\begin{figure}[H]
    \renewcommand{\figurename}{\textbf{Figure }}
    \renewcommand{\thefigure}{\textbf{1}}
     \centering
    \includegraphics[scale=0.6]{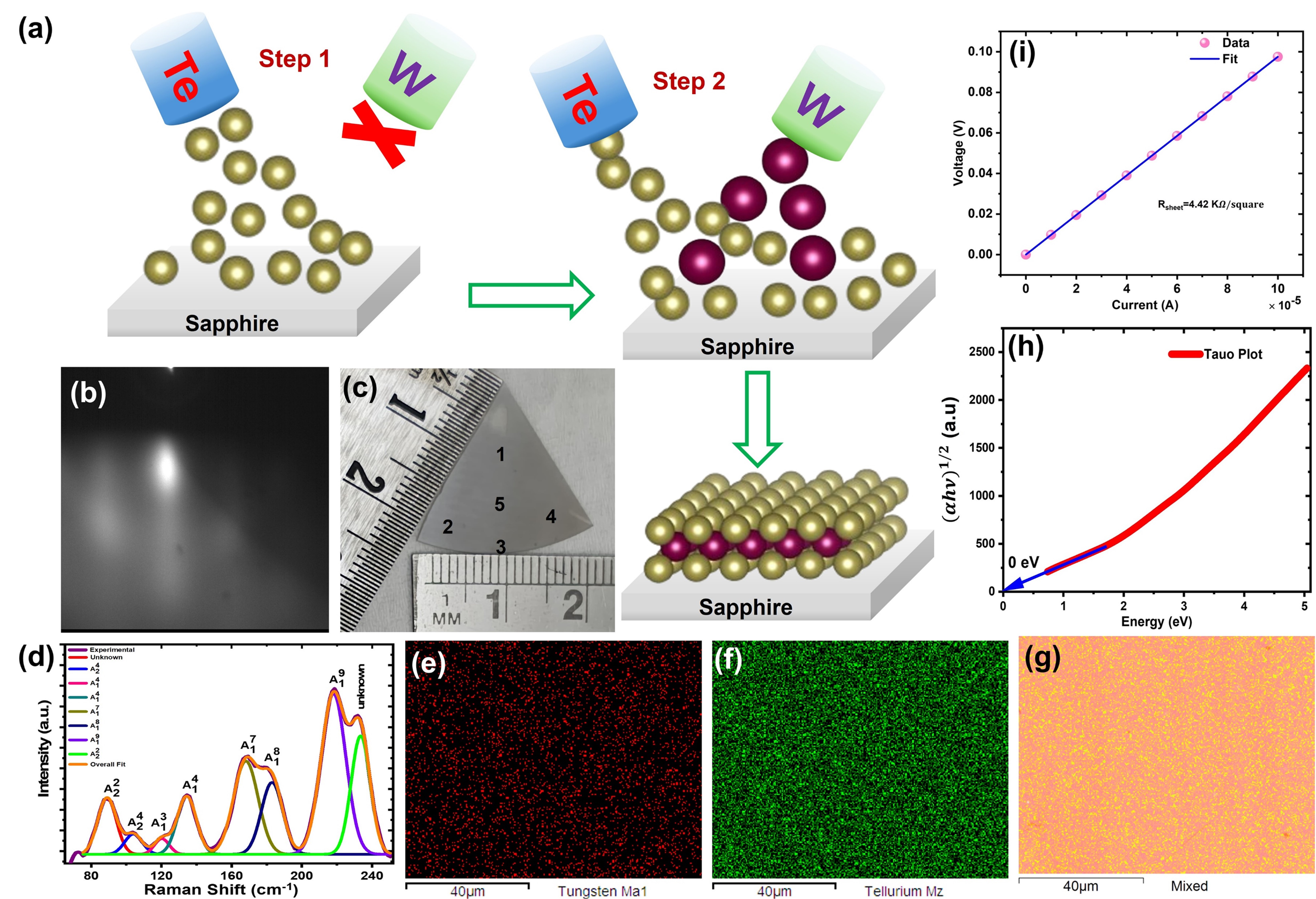}
    \caption{\textbf{Molecular Beam epitaxy Growth of1T$'$-WTe${2}$.} (a) Schematic of the pre-growth tellurisation of the substrate and consequent growth of 1T$'$-WTe${2}$ while opening both cracker cell of Tellurine and Tungsten  (b) Typical in-situ reflection high-energy electron diffraction (RHEED) image of grown flim. (c) Optical image of large-scale growth of 1T$'$-WTe${2}$ (d) Raman spectroscopy of grown film. (e-g) EDX mapping of tungsten, tellurium, and their combination at Position-5, as marked in figure 1c. (h) UV vis of  MBE grown 1T$'$-WTe${2}$. (i) Voltage vs current plot of MBE grown 1T$'$-WTe${2}$ using four-probe method. }
    \label{fig1}
\end{figure}

Figure-\ref{fig1}e shows the EDX map of tungsten, figure-\ref{fig1}f displays the EDX map of tellurium, and figure-\ref{fig1}g  presents the combined EDX mapping of both tungsten and tellurium at position-5. This shows consistent elemental composition, indicative of high-quality epitaxial growth in the central region. This serves as a representative analysis of the sample’s core properties. Meanwhile, the EDX maps for other positions (Positions 1–4) are provided in section-1 (S1) of the Supporting Information. Figure-\ref{fig1}h shows UV vis study of MBE grown 1T$^\prime$-WTe\textsubscript{2}. This gives us the insight of zero indirect optical bandgap. Figure-\ref{fig1}i shows the four probe electrical characteristic of MBE grown 1T$^\prime$-WTe\textsubscript{2}. In the four-probe setup, each probe was separated by 2~mm. For this configuration, the measured film resistance was approximately $R = 975.4\Omega$. The corresponding sheet resistance is given by $R_{\text{sheet}}$ = $\frac{\pi}{\ln 2} \cdot R$, where$\frac{\pi}{\ln 2} \approx 4.532$. Hence, $R_{\text{sheet}}$ $\approx$ 4.532 $\times$ 975.4 = $4420~\Omega/\square$. The sheet resistance is approximately $4.42~K\Omega/\square$. This proves that our grown 1T$^\prime$-WTe\textsubscript{2} film is suitable to be a van der Waals electrode for further electrical measurement.

\end{justify}

\section{\textbf{\large{Heterostructure characterisation and Device Fabrication:}}}

\subsection{\textbf{Atomic force microscopy (AFM) and X-ray Diffraction analysis (XRD):}}
\begin{justify}
The AFM image of MBE-grown 1T$^\prime$-WTe\textsubscript{2} is shown in figure-\ref{fig2}a and \ref{fig2}b. The roughness of this film is around 0.5 nm, which indicates good film quality and provides a solid foundation for the contact pad interface with InSe. The film thickness is 7.2 nm. Figures-\ref{fig2}c and \ref{fig2}d show the roughness and thickness of the InSe that we transferred on top of 1T$^\prime$-WTe\textsubscript{2} for our active channel. The roughness of InSe is around 0.28 nm, and the thickness of InSe on 1T$^\prime$-WTe\textsubscript{2} is 37.7 nm. The lattice parameters, phase purity, and interlayer distances can all be determined by analyzing the crystalline structure using X-ray diffraction (XRD). 

\begin{figure}[H]
        \renewcommand{\figurename}{\textbf{Figure }}
           \renewcommand{\thefigure}{\textbf{2}}
            \centering
	\includegraphics[scale=.53]{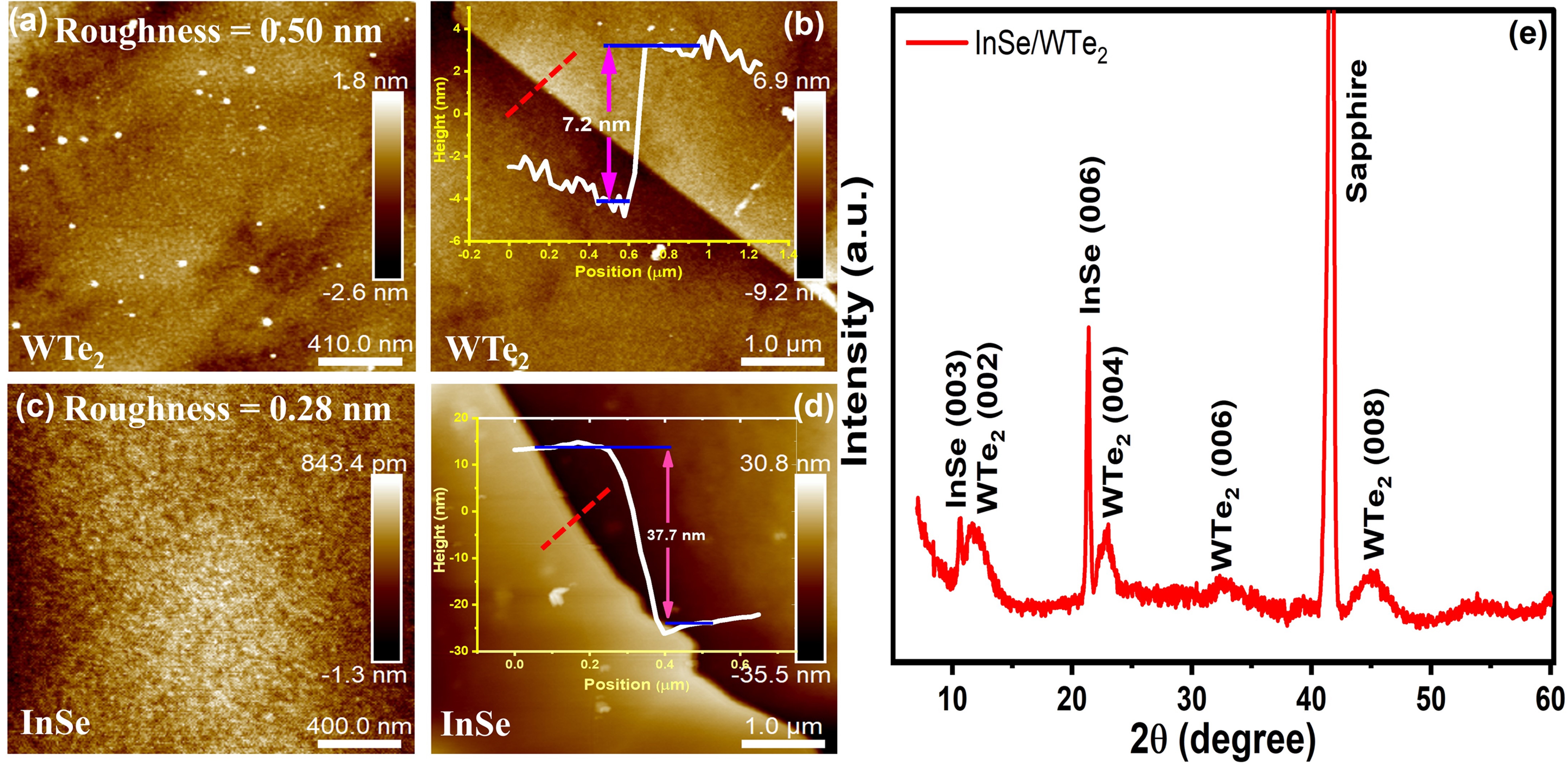}
		\caption{\textbf{AFM and XRD of InSe/ 1T$'$-WTe$_{2}$ :} (a) Roughness and (b) Thickness of MBE grown 1T$'$-WTe$_{2}$ pad. (c) Roughness and (d) Thickness of InSe. (e) XRD of InSe/1T$'$-WTe$_{2}$ configuration.} 
		\label{fig2}
\end{figure}
The orthorhombic crystal structure of the 1T$'$ phase of 1T$^\prime$-WTe\textsubscript{2} has a space group of Pmn2$_{1}$ \cite{brown1966crystal, augustin2000electronic}. This phase results in a dimerization of the W-W bonds, as tellurium (Te) atoms and tungsten (W) atoms form a deformed octahedral coordination. Weak van der Waals forces exist between layers of W and Te atoms stacked along the c-axis in the unit cell of 1T$^\prime$-WTe\textsubscript{2}. Prominent diffraction peaks corresponding to planes such as (002), (004), (006), etc., are generally seen in the XRD pattern of 1T$'$-WTe$_{2}$, indicating its layered structure. Figure-\ref{fig2} shows the XRD of InSe/1T$^\prime$-WTe\textsubscript{2} configuration. In this figure, there are several peaks For 1T$^\prime$-WTe\textsubscript{2} and InSe respectively. The interlayer spacing and lattice constants are represented by these peaks. For 1T$^\prime$-WTe\textsubscript{2}, the first peak occurs at an angle 2$\theta$ of 12.29$^{\circ}$, which corresponds to the (002) family of plane. From this, we calculated interlayer distance of 7.51\,\text{Å} and which corresponds to lattice constant of 14.4\,\text{Å} along the c-axis of 1T$^\prime$-WTe\textsubscript{2} \cite{brown1966crystal,augustin2000electronic, lee2015tungsten,zheng2016quantum}. Therefore, for this van der Waals material, the interlayer distance determined from the diffraction pattern is consistent with the expected values. Other typical peaks, such as (004), (006), and (008), were also observed, owing to the layered structure of 1T$'$-WTe$_{2}$. The well-ordered layered structure of 1T$^\prime$-WTe\textsubscript{2}, as shown by the XRD study, is characterized by peaks corresponding to distinct crystallographic planes. For Inse, we got first peak at an angle 2$\theta$ of 10.61$^{\circ}$ corresponding to (003) family of plane. and dominate peak at an angle 2$\theta$ of 21.32$^{\circ}$ corresponding to (006) family of plane. From this, we get d$_{003}$=8.33\,\text{Å} and corresponding c-axis parameter is approximately 16.66 \text{Å}, which is consistent with $\gamma$-phase of InSe \cite{weng2023polarization,li2022interfacial}.
\end{justify}

\subsection{\textbf{X-ray Photoelectron Spectroscopy (XPS) analysis :}}
\begin{justify}
The XPS analysis of 1T$^\prime$-WTe\textsubscript{2} and InSe, along with their heterostructure, provides crucial insights into their individual chemical compositions, bonding states, and interface properties. 1T$^\prime$-WTe\textsubscript{2} is a TMD that exhibits notable Weyl semimetal properties, contributing to its distinctive electrical characteristics. In XPS, the focus is on the core levels of tungsten (W) and tellurium (Te) of WTe$_{2}$, primarily represented by the W 4f and Te 3d peaks. The W 4f spectrum shows spin-orbit splitting, with the W 4f$_{7/2}$ peak typically appearing around 32 eV and the W 4f$_{5/2}$ around 34 eV, consistent with the expected values for WTe$_{2}$.

\begin{figure}[H]
        \renewcommand{\figurename}{\textbf{Figure }}
           \renewcommand{\thefigure}{\textbf{3}}
            \centering
	\includegraphics[scale=.8]{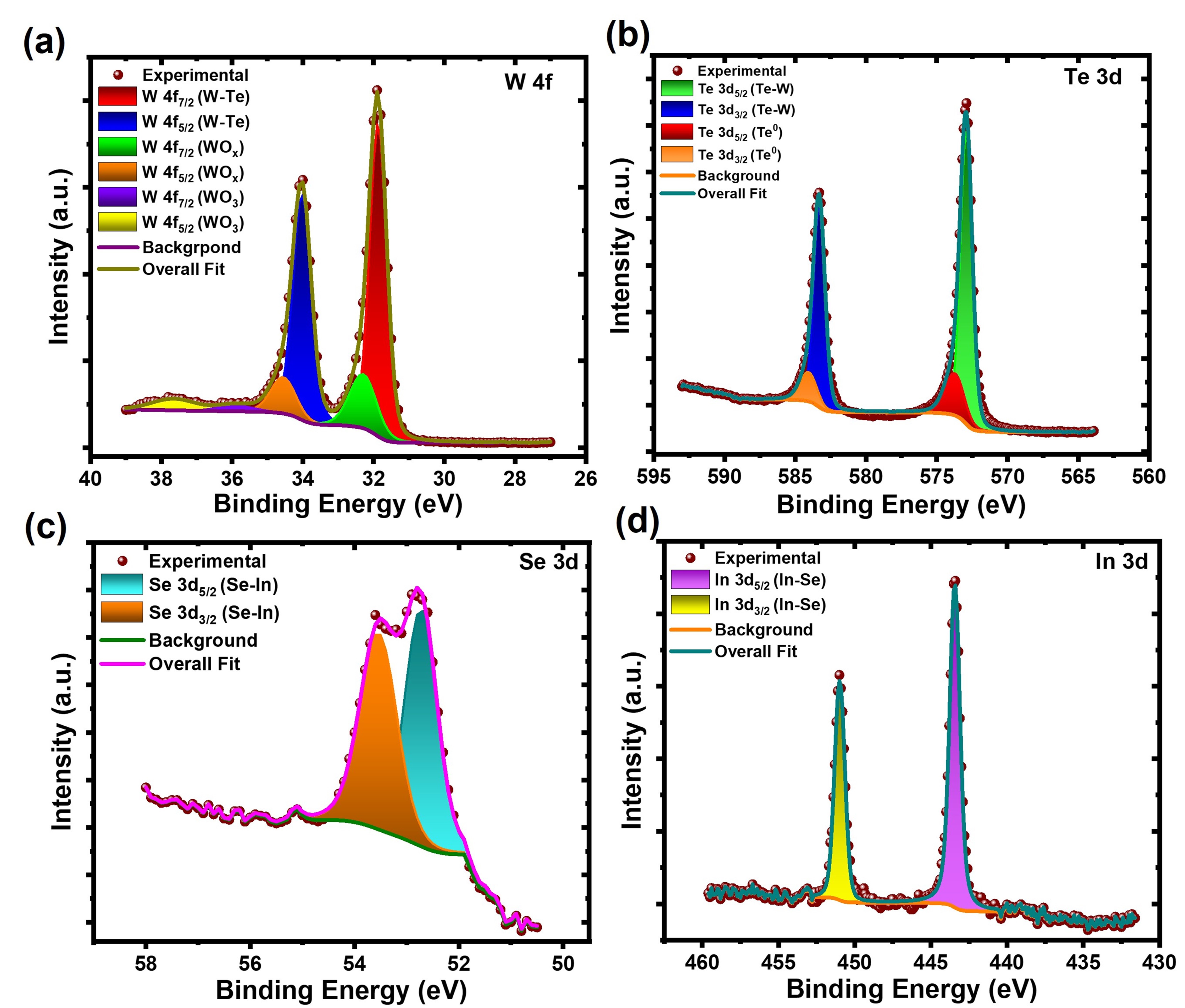}
		\caption{\textbf{XPS of InSe/1T$^{'}$-WTe$_{2}$ :} XPS analysis indicates that 1T$^{'}$-WTe$_{2}$ and InSe have a stoichiometric composition. (a) High resolution spectrum of 1T$^{'}$-WTe$_{2}$ for (a) W 4f core orbitals (b) Te 3d core orbitals. (c) High resolution spectrum of InSe for (a) Se 3d core orbitals (b) In 3d core orbitals.} 
		\label{fig3}
\end{figure}
The spin splitting energy between W 4f$_{7/2}$ and W 4f$_{5/2}$ is 2 eV. The tellurium 3d states are observed through the Te 3d peaks, with Te 3d$_{5/2}$ near 573 eV and Te 3d$_{3/2}$ at approximately 583 eV. The spin splitting energy between Te 3d$_{5/2}$ and Te 3d$_{3/2}$ is 10 eV \cite{mleczko2016high,chen2017simple}. InSe is a III-VI layered semiconductor known for its optoelectronic properties and a suitable bandgap for photonic applications. In XPS, the focus is primarily on the In 3d and Se 3d core levels of InSe. The In 3d$_{5/2}$ peak is typically observed at around 444.6 eV, with the In 3d$_{3/2}$ peak at 452.2 eV, showing a spin-orbit separation of about 7.6 eV. The Se 3d spectrum features two well-resolved peaks, with Se 3d$_{5/2}$ located at approximately 54.2 eV and Se 3d$_{3/2}$ at around 56.1 eV, separated by 0.9 eV. These peaks reflect the stable bonding of indium and selenium atoms within the material\cite{yang2017wafer}. Due to their complementary features, heterostructures made of WTe$_{2}$ and InSe are of great interest for applications in next-generation electrical and optoelectronic devices. XPS analysis of the 1T$^\prime$-WTe\textsubscript{2}/InSe structure allows for examination of the band alignment and interface chemistry between the two materials. The oxidation states of the atoms at the interface may change due to charge transfer between the WTe$_{2}$ and InSe layers. Changes in the chemical environment brought about by interlayer coupling can be identified by shifts in the binding energies of the W 4f, Te 3d, In 3d, and Se 3d peaks. Figure-\ref{fig2} shows XPS analysis of InSe/ 1T$'$-WTe$_{2}$ configuration. Figure-\ref{fig2}a shows high resolution spectrum of MBE grown 1T$'$-WTe$_{2}$ for W 4f core orbitals while figure-\ref{fig2}b shows for Te 3d core orbitals respectively after heterostructure configurarion with InSe. Figure-\ref{fig2}c shows high resolution spectrum of InSe for Se 3d core orbitals while figure-\ref{fig2}d shows for In 3d core orbitals respectively after heterostructure configurarion with 1T$^\prime$-WTe\textsubscript{2}. The W 4f spectrum shows two spin-orbit split orbitals: W 4f$_{7/2}$ at 31.89 eV and W 4f$_{5/2}$ at 33.98 eV, corresponding to the W-Te bond, with a spin splitting of 2.09 eV, comparable to individual 1T$^\prime$-WTe\textsubscript{2}. The shifts in W 4f$_{7/2}$ and W 4f$_{5/2}$ are 0.11 eV and 0.02 eV, respectively, due to the heterostructure configuration. Additionally, some oxide states due to charge transfer are also recorded at 32.30 eV (W 4f$_{7/2}$ for WO$_{x}$), 34.551 eV (W 4f$_{5/2}$, WO$_{x}$), and 32.39 eV (W 4f for WO$_{3}$). Similarly, the Te 3d of WTe$_{2}$ splits into two orbitals: Te 3d$_{5/2}$ at 572.92 eV and Te 3d$_{3/2}$ at 583.325 eV, with a separation of 10.405 eV. The shifts in Te 3d$_{5/2}$ and Te 3d$_{3/2}$ are 0.08 eV and 0.325 eV, respectively, due to the heterostructure configuration. Likewise, shifts in the binding energy of core levels for InSe are also observed in figure-\ref{fig3}. InSe's In 3d and Se 3d peaks are recorded in the XPS spectra, showing that in the heterostructure configuration, the In 3d splits into two spin-orbit peaks: In 3d$_{5/2}$ at 443.38 eV and In 3d$_{3/2}$ at 450.933 eV, with a spin splitting energy of 7.553 eV, which is comparable to individual InSe. The shifts in In 3d$_{5/2}$ and In 3d$_{3/2}$ are 1.22 eV and 1.2 eV, respectively, due to the heterostructure configuration. Similarly, Se 3d of InSe splits into Se 3d$_{5/2}$ at 52.731 eV and Se 3d$_{3/2}$ at 53.83 eV, with a spin splitting of 1.09 eV, comparable to the shift in individual InSe. This indicates that while the binding energy shifts due to the heterostructure configuration, the spin splitting energy remains unchanged. The stoichiometry of the 1T$^\prime$-WTe\textsubscript{2} film was determined from XPS by analyzing the core-level spectra of W 4f and Te 3d, using the CasaXPS software. The relative atomic concentrations were calculated based on the area under the fitted peaks, using the sensitivity factors provided in the CasaXPS elemental library. From this analysis, the W: Te ratio was found to be approximately 1:1.87, indicating a nearly stoichiometric WTe2 film, and for InSe, it is approximately 1:1.

\end{justify}

\subsection{\textbf{\large{Device fabrication of 1T$'$-WTe$_2$/InSe/1T$'$-WTe$_2$ and Ti-Au/InSe/Ti-Au:}}}
\begin{justify}

To construct a 1T$^\prime$-WTe\textsubscript{2}/InSe/1T$^\prime$-WTe\textsubscript{2}device, we initially patterned MBE-synthesized 1T$^\prime$-WTe\textsubscript{2} using Reactive Ion Etching (RIE). This step defines contact locations critical for device integration. A detailed description is provided in Section S2 of the Supporting Information. The final patterned structure is shown in figures S2c to S2d of Section S2 in the Supporting Information file. Following etching, a dry transfer technique was employed to accurately position InSe onto the patterned 1T$^\prime$-WTe\textsubscript{2} pads. A PDMS stamp affixed to a transparent glass slide was fabricated and cleaned. InSe flakes were mechanically exfoliated onto Nitto tape and subsequently transferred to the PDMS surface by controlled contact and peeling. The PDMS carrying InSe was aligned over the etched 1T$^\prime$-WTe\textsubscript{2} pads using a micromanipulator under an optical microscope. During the transfer, the substrate was gently heated to approximately 85\,\textdegree C to improve InSe adhesion via van der Waals interactions. The PDMS was then carefully lifted, leaving the InSe flake positioned on the 1T$^\prime$-WTe\textsubscript{2} contact region. This process is described in detail in Section S3 of the supporting Information. Figure S3g in the supplementary file shows the final device after etching and transfer. To fabricate a Ti–Au/InSe/Ti–Au device, we first exfoliate an InSe flake onto a sapphire substrate. Using our MicroTech laser writer, we perform lithography followed by metallization and lift-off to obtain the desired device structure. Optical images of the fabricated Ti–Au/InSe/Ti–Au device are shown in Figures S7a and S7b of the Supporting Information (Section S7). 
\end{justify}

\subsection{\textbf{\large{Barrier height estimation and Fermi level pinning:}}}
\begin{justify}
The contact potential difference (CPD) between the tip and the sample is given by \cite{maiti2020strain,melitz2011kelvin}
\begin{equation}
  V_{\text{CPD}} = \frac{\phi_{\text{tip}} - \phi_{\text{sample}}}{e} 
\end{equation}
Where \( V_{\text{CPD}} \) is the CPD, \( \phi_{\text{tip}} \) is the work function of the tip, \( \phi_{\text{sample}} \) is the work function of the sample, and \( e \) is the electronic charge.
The tip work function is determined to be 4.73 eV when using the standard HOPG substrate, with a contact potential difference (CPD) of 129 mV, as illustrated in figure-S5.1a of Section S5.1 of the Supporting Information. The CPD values corresponding to 1T$^\prime$-WTe\textsubscript{2} and InSe are 251 mV and 245 mV, respectively. Therefore, the work functions of 1T$^\prime$-WTe\textsubscript{2} and InSe are 4.478 eV and 4.484 eV, respectively. Next, we determine the Fermi level position of our InSe using XPS from the valence band spectra, as shown in figure-S5.1d of Section S5.1 of the Supporting Information. Using all of these data, we plot the disconnected band diagrams of the 1T$^\prime$-WTe\textsubscript{2}/InSe/1T$^\prime$-WTe\textsubscript{2} and InSe/Ti-Au devices in figure-S5.1e and figure-S5.1f, respectively, in Section S5.1 of the Supporting Information. Figure-S5.1g and figure-S5.1h of the Supporting Information show the equilibrium energy band diagrams of the 1T$^\prime$-WTe\textsubscript{2}/InSe/1T$^\prime$-WTe\textsubscript{2} and InSe/Ti-Au devices, respectively, in the ideal case, where Fermi level pinning is absent. The 1T$^\prime$-WTe\textsubscript{2}/InSe/1T$^\prime$-WTe\textsubscript{2} device exhibits a barrier height of 0.338 eV, while the InSe/Ti-Au device shows a barrier height of 0.96 eV at equilibrium, in the absence of pinning. This is one of the reasons why the 1T$^\prime$-WTe\textsubscript{2}/InSe/1T$^\prime$-WTe\textsubscript{2} device outperforms the InSe/Ti-Au-based device. In an ideal metal-semiconductor junction, the Schottky barrier height (SBH) for electrons is determined by the difference between the metal work function and the semiconductor electron affinity:

\begin{equation}
\phi_{Bn,\text{ideal}} = \phi_m - \chi
\label{eq:ideal_sbh}
\end{equation}

This is known as the Schottky-Mott limit and is valid under ideal, unpinned conditions. However, in practical cases, deviations from this ideal behavior occur due to the formation of interface states and metal-induced gap states (MIGS), which pin the Fermi level near the charge neutrality level (CNL). This phenomenon is referred to as Fermi level pinning. Under such pinning conditions, the barrier height becomes nearly independent of the metal work function. Depending on the pinning factor \( S \),different scenarios arise. The Fermi level is pinned exactly at the charge neutrality level, i.e.,  S = 0. This is called strong pinning. For an ideal unpinned condition, the Fermi level in a semiconductor in a metal-semiconductor contact depends only on the metal work function, i.e., S = 1. In real experimental scenarios, the pinning factor typically lies between these two extremes, $0 \leq S \leq 1$. In practice, the experimental Schottky barrier height is given by \cite{kim2017fermi}
\begin{equation}
\phi_{Bn,exp} = S(\phi_m - \phi_{\text{CNL}}) + \phi_{\text{CNL}} - \chi \approx E_F-\chi \approx E_F-E_C
\label{eq:experimental_sbh}
\end{equation}
where $\phi_{Bn,\text{exp}}$ is the experimental barrier height, which includes the effects of Fermi level pinning, \( \phi_{\text{CNL}} \) is the charge neutrality level and \( \chi \) is the electron affinity of the semiconductor. To extract the barrier height experimentally, including the effects of Fermi level pinning in the devices, we measured the current as a function of applied voltage at different temperatures. Figures-\ref{fig4}a and \ref{fig4}b show the current-voltage (I-V) characteristics of 1T$^\prime$-WTe\textsubscript{2}/InSe/1T$^\prime$-WTe\textsubscript{2} and InSe/Ti-Au, respectively, measured at 300K, 351K, 385K, and 436K. I-V plots for the remaining temperatures are provided in figures-S4a and S4b in Supplementary Section S4. To extract the barrier height from our experimentally measured data, we employed Richardson-Dushman model\cite{jeon2018epitaxial} 
\begin{equation}
I = AA^* T^2 \exp\left( -\frac{q\phi_{B,\text{exp}}}{k_B T} \right) \left[ \exp\left( \frac{qV}{n k_B T} \right) - 1 \right]
\label{eq:richardson_dushman}
\end{equation}

where $A$ is the contact area, $A^*$ is the Richardson constant, $T$  is the absolute temperature, $q$ is the elementary charge, $\phi_{B,\text{exp}}$ is the experimental Schottky barrier height, $k_B$ is Boltzmann's constant, $n$ is the ideality factor, and $V$ is the applied voltage. For sufficiently forward bias, where $V \gg \frac{k_B T}{q}$, the current becomes approximately exponential:
\begin{equation}
I \approx AA^* T^2 \exp\left( -\frac{q\phi_{B,\mathrm{exp}}}{k_B T} \right) \exp\left( \frac{qV}{n k_B T} \right) = I_0 \exp\left( \frac{qV}{n k_B T} \right)
\end{equation}
\begin{figure}[!t]
        \renewcommand{\figurename}{\textbf{Figure }}
           \renewcommand{\thefigure}{\textbf{4}}
            \centering
	\includegraphics[scale=1]{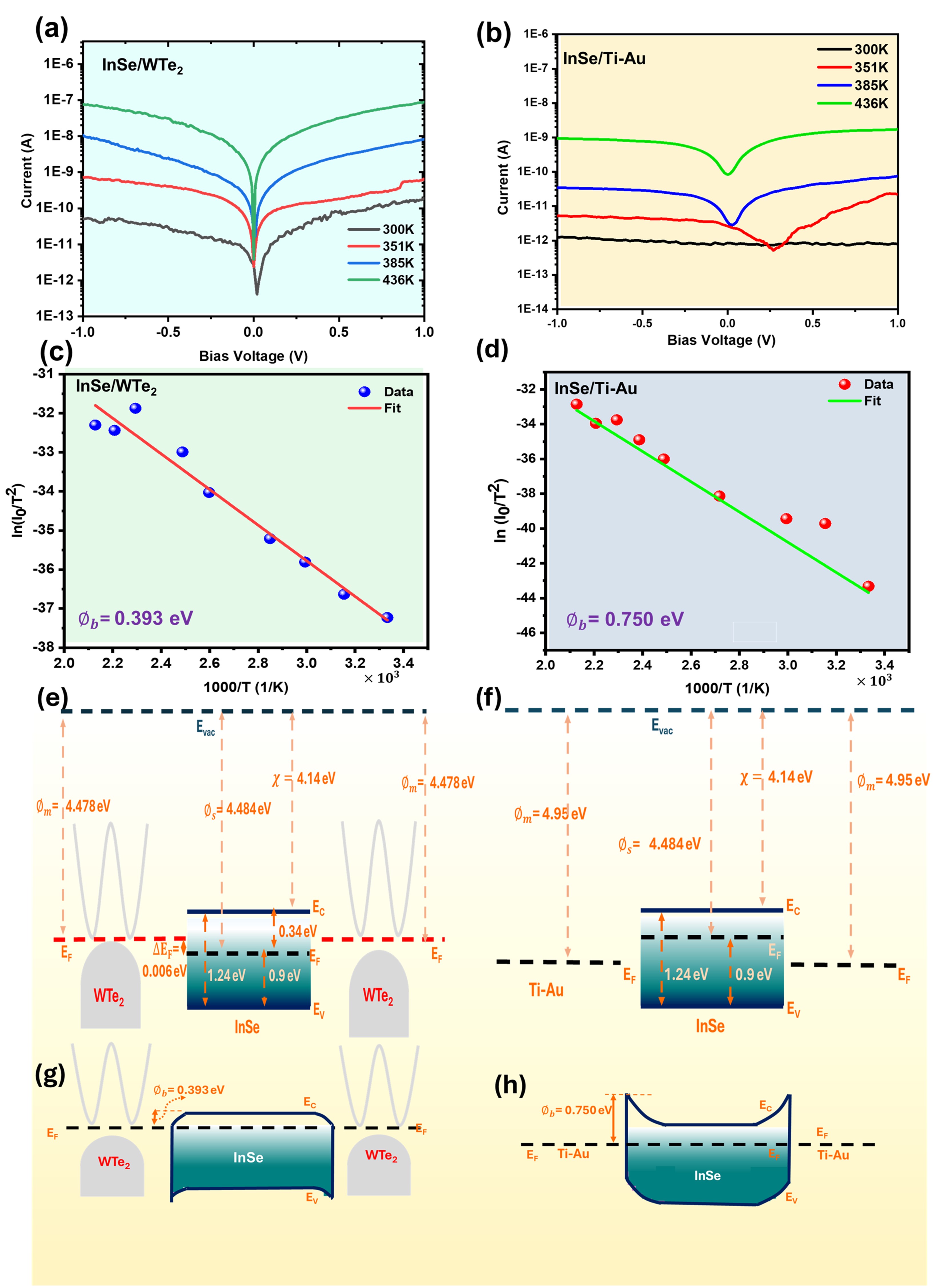}
		\caption{\textbf{Current versus voltage characteristics and Band diagram of 1T$^\prime$-WTe\textsubscript{2}/InSe and InSe/Ti-Au based device :} Current versus Voltage characteristics of (a) 1T$^\prime$-WTe\textsubscript{2}/InSe (b) InSe/Ti-Au devices at different temperatures, 300K, 351K, 385K and 436K, respectively. Richardson plot of (c) 1T$^\prime$-WTe\textsubscript{2}/InSe device and (d) InSe/Ti-Au device respectively. Band diagram of (e) 1T$^\prime$-WTe\textsubscript{2}/InSe device and (f) InSe/Ti-Au, before connection respectively. Energy band diagram of (g)1T$^\prime$-WTe\textsubscript{2}/InSe device and (h) InSe/Ti-Au device under equilibrium, respectively, after accounting for the pinning factor.}
		\label{fig4}
\end{figure}

Where,
\[
I_0 = AA^* T^2 \exp\left( -\frac{q\phi_{B,\mathrm{exp}}}{k_B T} \right)
\]

is the saturation current. Therefore, the experimental barrier height $\phi_{\mathrm{B,exp}}$ can be extracted from the plot of $\ln(I_0/T^2)$ versus $1/T$, known as the Richardson plot. To do this, we focus on the forward bias regime ($V \gg \frac{k_B T}{q}$) of the measured $I$–$V$ data. In this regime, we plot $\ln(I)$ vs. $V$ at each temperature and extract $I_0$ by fitting the data. These fits are shown in Supplementary Section S4. Specifically, figure-S4(b) and figure-S4(d) display the $\ln(I)$ vs. $V$ curves for the InSe/WTe\textsubscript{2} and InSe/Ti-Au devices, respectively, at different temperatures. From these curves, we obtain $I_0$ at each temperature for both device types. We then plot $\ln(I_0/T^2)$ versus $1000/T$ to generate the Richardson plots, from which we extract the experimental barrier height $\phi_{\mathrm{B,exp}}$ from the slope. Figure-\ref{fig4}(c) and figure-\ref{fig4}(d) show the Richardson plots for the InSe/1T$^\prime$-WTe\textsubscript{2} and InSe/Ti-Au devices, respectively. From these plots, we find $\phi_{\mathrm{B,exp}}^{\mathrm{InSe/WTe_2}} = 0.393$~eV {$\phi_{\mathrm{B,exp}}^{\mathrm{InSe/Ti\text{-}Au}} = 0.750$~eV. Clearly, the barrier height of the InSe/Ti-Au device is approximately twice that of the 1T$^\prime$-WTe\textsubscript{2}/InSe device. Figure-\ref{fig4}e and figure-\ref{fig4}f show the band diagrams of 1T}$^\prime$-WTe\textsubscript{2}/InSe/1T$^\prime$-WTe\textsubscript{2} and Ti-Au/InSe/Ti-Au before connection, respectively. Figure-\ref{fig4}g and figure-\ref{fig4}h show the same after connection, where the pinning factor is included.
\end{justify}

\subsection{\textbf{\large{Experimental extraction of contact resistance:}}}
\begin{justify}
To precisely assess the contact resistance between 1T$'$-WTe$_2$/InSe and Ti--Au/InSe heterointerfaces, we employed the four-probe Transmission Line Method (TLM) \cite{schroder2015semiconductor}. The four-probe TLM is a well-established and highly accurate characterization technique for directly extracting the intrinsic contact resistance in semiconductor–metal junctions. By eliminating parasitic effects such as probe resistance and probe–to–contact resistance, the method ensures that the measured data reliably reflect the true contact and sheet resistance of the system under investigation. Moreover, the four-probe TLM approach effectively mitigates anomalies, including voltage drop inaccuracies, series resistance contributions, and current crowding effects, which often compromise two-probe measurements \cite{schroder2015semiconductor}. As a result, electrical transport characterization in layered materials and heterostructures becomes more accurate and reliable in this four-probe TLM method. For this purpose, the MBE-grown 1T$'$-WTe$_2$ is first patterned into a TLM structure using lithography, followed by reactive ion etching (RIE). Similarly, the Ti-Au TLM structures are fabricated through a standard lithography process, followed by metal deposition and lift-off. Figures S6.1a and S6.2a show the TLM structures of 1T$'$-WTe\textsubscript{2} and Ti–Au, respectively, as presented in Section S6 of the Supporting Information. Next, we transfer InSe onto the newly fabricated TLM structures using our 2D transfer system. Figure-\ref{fig5}a and \ref{fig5}b show the TLM structure of 1T$'$-WTe\textsubscript{2}/InSe, while Figure-\ref{fig5}c and \ref{fig5}d present the TLM structure of Ti-Au/InSe. Figure-\ref{fig5}e illustrates the measurement setup for performing four-probe TLM measurements. For this purpose, we use a Keithley 2450 Source Measure Unit (SMU) with Python-based automation to source, record, and store the data in Excel format. To conduct the TLM measurements, we vary the channel length ($L$) between 3~µm, 4~µm, and 5~µm. In the four-probe configuration, the forcing terminals of the Keithley 2450 SMU are connected to the outer pads, as shown in Figure-\ref{fig5}e, while the sensing terminals are connected to the inner pads, which are separated by 3~µm. It should be noted that for the $L = 3$~µm measurement, the forcing pads are also 3~µm apart from the sensing pads. A current sweep is applied through the forcing terminals, and the corresponding voltage is recorded using the Keithley 2450 SMU with automation for $L = 3$~µm. The same procedure is repeated for $L = 4$~µm and $L = 5$~µm, respectively. Figure S6.1 (e)-(g) of Supporting Information of section S6, shows the measured voltage versus forcing current characteristics for $L = 3~\mu$m, $4~\mu$m, and $5~\mu$m, respectively, obtained using the four-probe method for the 1T$'$-WTe\textsubscript{2}/InSe TLM structure. From these plots, the resistance values for $L = 3~\mu$m, $4~\mu$m, and $5~\mu$m were extracted by fitting only the linear region. Since the sample and contacts follow Ohm's law in this regime, only the low-current linear region can be fitted in four-probe TLM measurements. The voltage–current relationship is linear at low currents and represents the intrinsic resistance of the material and contacts. It is important to note that the device width varies for $L = 3~\mu$m, $4~\mu$m, and $5~\mu$m. To obtain the correct resistance, the measured resistance must therefore be normalized with respect to the device width. 
\begin{figure}[!t]
        \renewcommand{\figurename}{\textbf{Figure }}
           \renewcommand{\thefigure}{\textbf{5}}
           \centering
	\includegraphics[scale=1]{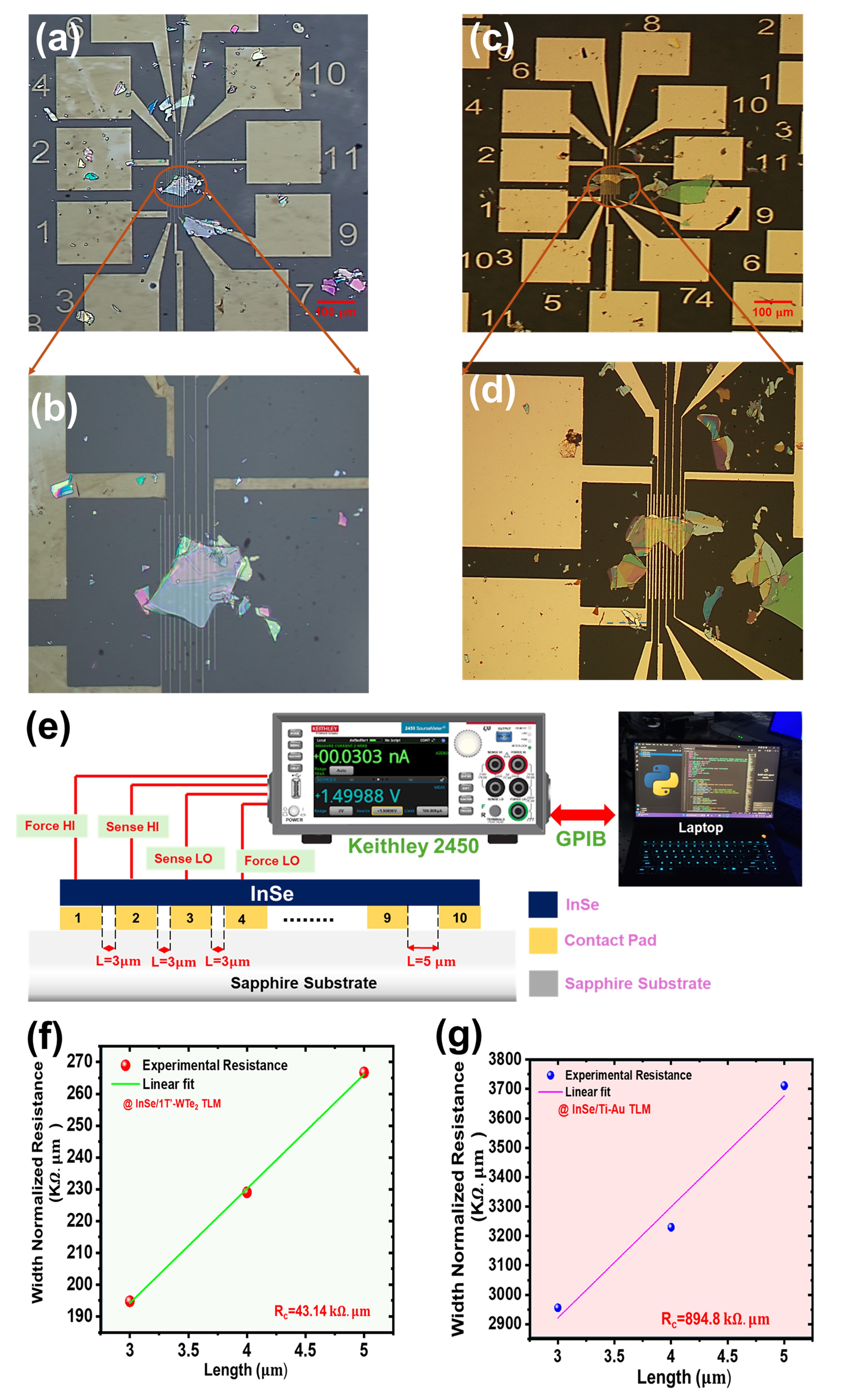}
		\caption{\textbf{TLM structure of InSe/1T$'$-WTe\textsubscript{2} and InSe/Ti–Au devices and their contact resistance measurement:} 
(a) TLM structure of 1T$'$-WTe\textsubscript{2}/InSe. 
(b) Enlarged view of 1T$'$-WTe\textsubscript{2}/InSe TLM. 
(c) TLM structure of Ti–Au/InSe. 
(d) Enlarged view of Ti–Au/InSe TLM. 
(e) Four-probe TLM measurement setup using a Keithley 2450 SMU with Python automation. 
(f) Width-normalized resistance versus channel length for InSe/1T$'$-WTe\textsubscript{2}. 
(g) Width-normalized resistance versus channel length for InSe/Ti–Au devices, respectively.}

\label{fig5}
\end{figure}
Figure S6.1 (d) of Supporting Information of section S6 presents an enlarged optical image of the 1T$'$-WTe\textsubscript{2}/InSe TLM structure, where the width at each pad is indicated in red colour. For instance, the width at pad~2 is $35.774~\mu$m, while at pad~3 it is $43.081~\mu$m. For the $L = 3~\mu$m four-probe measurement, the sensing terminals are pad~2 and pad~3, as shown in Figure-\ref{fig5}e. Thus, the effective width for $L = 3~\mu$m is calculated by averaging the widths at pads 2 and 3, yielding $39.43~\mu$m. Similarly, the effective widths for $L = 4~\mu$m and $L = 5~\mu$m are $76.58~\mu$m and $81.02~\mu$m, respectively. After normalizing the resistance by the corresponding device width, we plot the width-normalized resistance as a function of channel length ($L$), as shown in Figure-\ref{fig5}f. Similarly, for the Ti-Au/Inse TLM structure, figure S6.2 (e)-(g) of the Supporting Information of section S6, shows the measured voltage versus forcing current characteristics for $L = 3~\mu$m, $4~\mu$m, and $5~\mu$m, respectively, obtained using the four-probe method. Figure S6.2 (d) of the Supporting Information of section S6 presents an enlarged optical image of the Ti-Au/InSe TLM structure, where the width at each pad is indicated in red colour. In the same process, we get the effective width of the device for each channel length and again, we plot the width-normalized resistance as a function of channel length ($L$), as shown in Figure-\ref{fig5}g for Ti-Au/InSe TLM structure. The measured total resistance can be expressed as sum of twice of contact resistance (as there are two contacts) and channel resistance \cite{schroder2015semiconductor}. Therefore, 
\begin{equation}
R_{\text{Total}}(L) = 2R_{C} + \frac{R_{\text{sh}}}{W}L
\end{equation}
Where, $R_{\text{Total}}(L)$ is the measured resistance for different lengths obtained using the four-probe method.  
$R_C =$ Contact resistance  
$R_{sh} =$ Sheet resistance  
$W =$ Width of the device  
$L =$ Contact spacing or device length

The width-normalized resistance is expressed as:  

\begin{equation}
R_{\text{Nor},W}(L) = R_{\text{Total}}(L) \cdot W = 2R_C W + R_{sh} L
\end{equation}

From this relation, the contact resistance is obtained by extracting the $y$-intercept of the fitted curve and dividing it by 2, with $W = 1\mu$m. Hence, the extracted contact resistance represents the contact resistance per unit width (per $\mu$m). By fitting the data in figure~\ref{fig5}g and figure~\ref{fig5}f, the $y$-intercepts for the 1T$'$-WTe\textsubscript{2}/InSe and Ti–Au/InSe structures are obtained, from which their respective contact resistances are calculated. The extracted contact resistance of the WTe\textsubscript{2}/InSe TLM structure is $R_{C,\text{WTe}_2/\text{InSe}} = 43.14~\text{k}\Omega \cdot \mu\text{m}$. Following the same procedure for the Ti–Au/InSe TLM structure (figure-\ref{fig5}g), the contact resistance is found to be $R_{C,\text{Ti–Au}/\text{InSe}} = 894.8\text{k}\Omega \cdot \mu\text{m}$. These results clearly indicate that the 1T$'$-WTe$_2$/InSe device exhibits approximately 21 times lower contact resistance and nearly half the barrier height compared to the Ti-Au/InSe device. This significant improvement highlights the effectiveness of using semimetallic 1T$'$-WTe$_2$ contacts for achieving superior charge injection, reduced interface scattering, and improved overall device performance.
\end{justify}

\section{\textbf{\large{Experimental Photoresponse measurements:}}}
\subsection{\textbf{\large{Optoelectronic Response Analysis:}}}
\begin{justify}
    To evaluate the performance of 1T$^\prime$-WTe\textsubscript{2} as a 2D van der Waals electrode in a 2D layered material system, we fabricated a 2D/2D contact-based photodetector. In this configuration, InSe serves as the channel material, while the semimetallic 1T$^\prime$-WTe\textsubscript{2} functions as the van der Waals electrode. InSe was chosen as the channel material due to its broad-spectrum responsivity, allowing us to clearly observe the effects of the proposed van der Waals electrode made from 1T$^\prime$-WTe\textsubscript{2} across a wide range of wavelengths. Figure-\ref{fig5}a shows SEM image and figure-\ref{fig5}b depicts the schematic of the 1T$^\prime$-WTe\textsubscript{2}/InSe photodetector. To compare the device performance, we also fabricated an InSe-based photodetector on a sapphire substrate using conventional metal contacts, referred to as the InSe/Ti-Au photodetector. The area of 1T$^\prime$-WTe\textsubscript{2}/InSe and InSe/Ti-Au devices are 200 $\mu \text{m}^2$ and 30 $\mu \text{m}^2$ approximately. The thickness of the InSe in the InSe/Ti-Au device is 30.6 nm, which is comparable to the thickness of InSe (37.7 nm) in the 1T$^\prime$-WTe\textsubscript{2}/InSe device. The device structure and optical image of this InSe/Ti-Au photodetector are shown in Figure S6 of the Supporting Information. The devices were measured in a two-terminal configuration in ambient conditions at room temperature. Figure-\ref{fig5}c and \ref{fig5}d show the current vs bias voltage characteristics of a photodetector in the presence and without light, based on 1T$^\prime$-WTe\textsubscript{2}/InSe and InSe/Ti-Au configurations, respectively. Here, we expose both devices with different wavelength from 250 nm (UV), 630 nm (visible) and 950 nm (NIR) with power 0.332 $\mu \text{W/mm}^2$, 117.7 $\mu \text{W/mm}^2$ and 267 $\mu \text{W/mm}^2$ respectively. The current vs bias voltage characteristics of a photodetector in the presence (for wavelengths 200 nm to 1100 nm ) and absence of light, based on 1T$^\prime$-WTe\textsubscript{2}/InSe and InSe/Ti-Au configurations, are shown in figures S6d and S6f, respectively, in section S6 of the supporting information. Extra electron-hole pairs were generated through the band-to-band transition when the film absorbed the light with energy higher than the band gap of InSe and then separated by the drain-source bias. 
    
\begin{figure}[!t]
        \renewcommand{\figurename}{\textbf{Figure }}
           \renewcommand{\thefigure}{\textbf{6}}
           \centering
	\includegraphics[scale=.7]{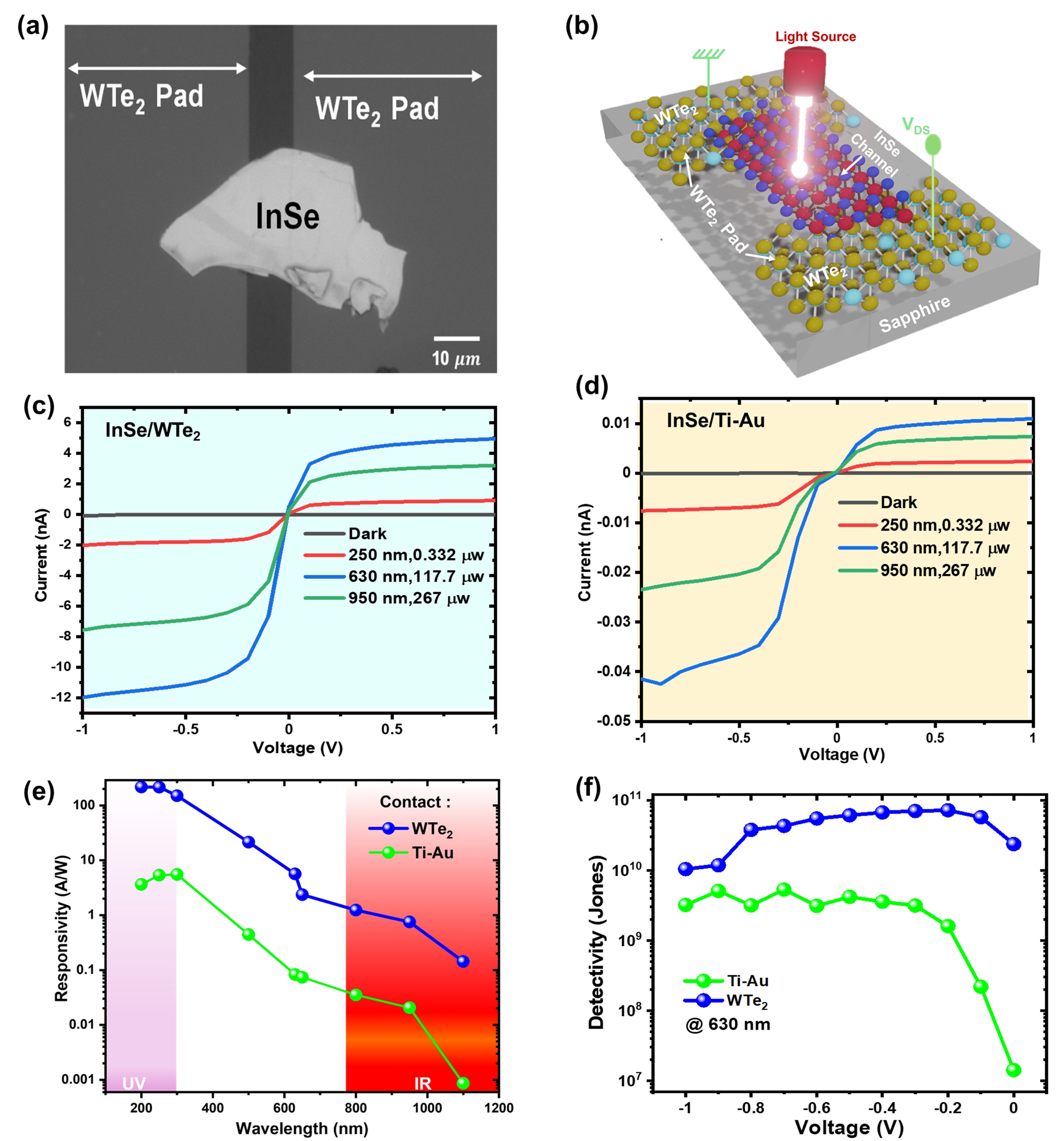}
		\caption{\textbf{Current Voltage (I-V) characteristics, responsivity and detectivity of InSe/ 1T$'$-WTe$_{2}$ and InSe/ Metal devices:} (a) Scanning electron microscope (SEM) image  (b) schematic of InSe/ 1T$'$-WTe$_{2}$ device. I-V characteristics at dark, 250 nm, 630 nm and 950 nm light conditions for (c) InSe/1T$'$-WTe$_{2}$ device. and (d) InSe/Ti-Au device. (e) Responsivity and (f) detectivity of InSe device with 1T$'$-WTe$_{2}$ contact (blue) and metal contact (green) respectively. } 
		\label{fig6}
\end{figure}
Then the photogenerated free electrons and holes transported toward the electrodes in opposite directions, which increase the drain current as the photocurrent $(I_{\text{ph}} = I_{\text{light}} - I_{\text{dark}})$, where \( I_{\text{light}} \) represents the current in the presence of light, \( I_{\text{dark}} \) denotes the current in darkness. The performance of a phototransistor can be comprehensively assessed through three primary figures of merit: photoresponsivity, external quantum efficiency, and response time. The responsivity, denoted as \( R_{\lambda} \), is calculated by taking the ratio of the photocurrent (\( I_{\text{ph}} \)) to the product of the incident power density (\( P \)) and the area that is illuminated (\( A \)). Hence, photoresponsivity (\(R_{\lambda}\)) quantifies the efficiency of a photodetector when exposed to incident photons. It is defined as the ratio of the photogenerated current to the power of the incoming light incident on the effective area of the device. \begin{equation}
    R_{\lambda} = \frac{I_{\text{ph}} }{P_{\lambda} A},
\end{equation}
where, Where \( I_{\text{ph}} \) represents the photo current ,\( P_{\lambda} \) is the power density of the incident light, and \( A \) refers to the effective area of the device.This relationship allows for a clear understanding of how effectively the phototransistor converts incident light into electrical current, providing critical insights into its overall functionality and efficiency. In evaluating photodetector efficiency, the Photo-Detectivity Coverage Ratio (PDCR) is a key metric that assesses a photodetector’s ability to distinguish signals from noise under specific conditions. PDCR measures how effectively a photodetector converts incident light into an electrical signal by comparing the photocurrent (current generated in response to light) with the dark current (current in the absence of light). A higher PDCR indicates enhanced performance, as it suggests the detector is better at differentiating light signals from background noise. PDCR is essential for determining the sensitivity of photodetectors. Higher values imply that the detector can more accurately separate light signals from noise, which is especially useful in applications requiring high precision, such as imaging and optical sensing. As a result, PDCR is a critical parameter for evaluating and comparing photodetectors' sensitivity and effectiveness, particularly in low-light or precision-demanding environments. PDCR can be expressed as\cite{khan2024unveiling}:
 \begin{equation}
   \text{PDCR} = \frac{I_{\text{Light}} - I_{\text{dark}}}{I_{\text{dark}}}
\end{equation}

Detectivity (\(D\)) indicates the capability of a photodetector to identify weak signals, and it is quantitatively measured using the following expression.

 \begin{equation}
    D = \frac{R_{\lambda} \sqrt{A}}{\sqrt{2 e I_{\text{dark}}}}.
\end{equation}

 To gain insight into the impact of 1T$'$-WTe$_{2}$ as a van der Waals electrode on the optoelectronic properties of the device, we plotted spectral response ie the responsivity as a function of wavelength, spanning from deep ultraviolet (DUV) to near-infrared (NIR), for both the 1T$^\prime$-WTe\textsubscript{2}/InSe and InSe/Ti-Au configurations, as illustrated in figure-\ref{fig5}e. Our analysis revealed a remarkable photoresponsivity of 217.58 A/W in the DUV region for the 1T$^\prime$-WTe\textsubscript{2}/InSe configuration, whereas the InSe/Ti-Au setup exhibited a significantly lower photoresponsivity of 3.64 A/W in the same region. In the visible spectrum, particularly at a wavelength of 630 nm, the photoresponsivity for 1T$^\prime$-WTe\textsubscript{2}/InSe reached 5.66 A/W, compared to 0.08 A/W for the InSe/Ti-Au configuration. Similarly, in the NIR range at 950 nm, the InSe1T$'$-WTe$_{2}$ demonstrated a photoresponsivity of 0.143 A/W, while the InSe/Ti-Au configuration showed a considerably lower value of $8.65 \times 10^{-4}$ A/W. Overall, across the entire spectrum from DUV to NIR, it is evident that the InSe/1T$'$-WTe$_{2}$ device significantly outperforms its InSe/Ti-Au counterpart, highlighting the superior optoelectronic performance imparted by the van der Waals electrode. In the context of PDCR, for 630 nm light, for InSe/1T$'$-WTe$_{2}$ and InSe/Ti-Au devices, $I_{Light}$ are 1.27 $\times 10^{-8}$ A and 4.02 $\times 10^{-11}$ A, respectively. The dark current, $I_{Dark}$ for InSe/1T$'$-WTe$_{2}$ and InSe/Ti-Au devices are 1.92 $ \times 10^{-12}$ A and 1.63  $\times 10^{-14}$ A respectively. Hence, PDCR at 630 nm for 1T$^\prime$-WTe\textsubscript{2}/InSe and InSe/Ti-Au devices are 6613 and 2465 respectively. This makes 1T$^\prime$-WTe\textsubscript{2}/InSe device 2.7 times more sensitive compared to the InSe/Ti-Au device. Figure-\ref{fig5}f shows the ditectivity of 1T$^\prime$-WTe\textsubscript{2}/InSe (blue) and InSe/Ti-Au (green) devices at 630 nm. Here also, 1T$^\prime$-WTe\textsubscript{2}/InSe outperforms the InSe/Ti-Au device.
 
\end{justify}
\subsection{\textbf{\large{Response Time analysis:}}}
\begin{justify}
The response time is a critical parameter for evaluating the transient behavior of a photodetector. Response time measures how quickly the photodetector reacts to changes in the modulating signal. It is assessed on both the rising (rise time) and falling (fall time) edges of the signal. To determine the response time, the fast $\tau_{1}$ and slow $\tau_{1}$ components of the rise and fall times are extracted. This approach involves fitting the experimental data to a bi-exponential equation \cite{kaushik20212d,kandar2025scalable}. 
\begin{equation}
I(t)=I_{0} + A_1 e^{-(t-t_{0})/\tau_{1}} + A_2 e^{-(t-t_{0})/\tau_{2}}
\label{equ 9}
\end{equation}
where, $I_{0}$ is the steady-state current at $t = t_{0}$, and $\tau_{1}$ and $\tau_{2}$ represent the fast rise (or fast decay) time and the slow rise (or slow decay) time, respectively. For the rising edge of the photocurrent, we denote $\tau_{r1}$ as the fast rise time and $\tau_{r2}$ as the slow rise time. The slow rise time, $\tau_{r2}$, indicates that the photodetector takes longer to reach its peak response after light exposure begins. This delay may occur due to charge trapping and recombination, where charge carriers become temporarily trapped at defect sites before contributing to the overall current. Hence, this phenomenon may be referred to as charge trapping rise delay (CTRD), as it is associated with protracted processes such as charge trapping and delayed carrier recombination. Similarly, for the falling edge of the photocurrent, we denote $\tau_{d1}$ as the fast decay time and $\tau_{d2}$ as the slow decay time. The slow decay time, $\tau_{d2}$, indicates that the photodetector takes longer to return to its dark level after the light is turned off. This delay may also result from charge trapping and recombination, with charge carriers gradually releasing from defect sites over time, contributing to the extended decay.

\begin{figure}[!t]
        \renewcommand{\figurename}{\textbf{Figure }}
           \renewcommand{\thefigure}{\textbf{7}}
           \centering
	\includegraphics[scale=0.87]{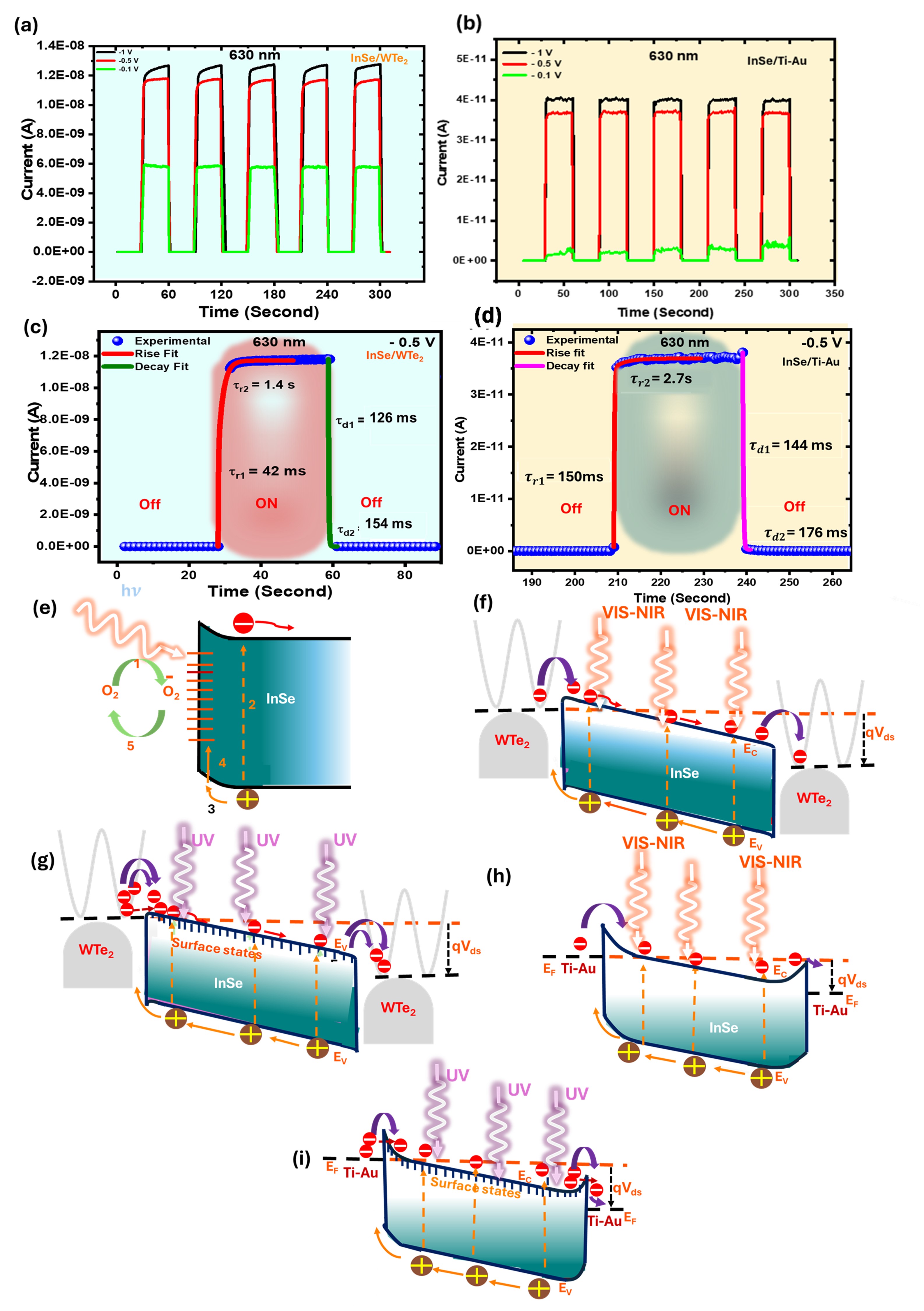}
		\caption{\textbf{Current Vs time (I-t)  characteristics of InSe/ 1T$'$-WTe$_{2}$ and InSe/Ti-Au devices @ 630 nm:} (a) I-t characteristics at different bias voltages: -1V, -0.5V and -0.1V respectively InSe/ 1T$'$-WTe$_{2}$ (b) Enlarged I-t charcteristics of InSe/ 1T$'$-WTe$_{2}$ at -0.5 V for the for the fitting. The red curve shows a rise fit, while the green curve shows a decay fit, respectively, for the InSe/1T$'$-WTe$_{2}$ device. (c) I-t characteristics at different bias voltages: -1V, -0.5V, and -0.1V, respectively. InSe/ Metal (d) Time resolved I-t characteristics of InSe/Ti-Au at -0.5 V for the fitting. The red curve shows a rise fit, while the pink curve shows a decay fit, respectively, for the InSe/Ti-Au device. (e) Schematic diagram of different steps in oxygen-sensitized photoconduction (OSPC). Energy band diagram of InSe/ 1T$'$-WTe$_{2}$ device under drain to source bias when it is exposed to (f) Visible (VIS) and near infrared (NIR) light and (g) Deep Ultraviolet light (DUV), respectively.  Energy band diagram of InSe/ Metal device under drain to source bias when it is exposed to (h) Visible (VIS) and near infrared (NIR) light, and (i) Deep Ultraviolet light (DUV), respectively.} 
		\label{fig6}
\end{figure}

To further investigate the influence of 1T$'$-WTe$_{2}$ as a van der Waals electrode, we compare the time response characteristics of both InSe/1T$'$-WTe$_{2}$ and InSe/Ti-Au devices. As illustrated in figure-\ref{fig6}a, the time response for the InSe/1T$'$-WTe$_{2}$ configuration is presented. Meanwhile, figure-\ref{fig6}c depicts the fitted time response for InSe/1T$'$-WTe$_{2}$ at a wavelength of 630 nm. In a similar manner, figure-\ref{fig6}b shows the time response of the InSe/Ti-Au device, and figure-\ref{fig6}d provides the fitted time response for the InSe/Ti-Au configuration at 630 nm. These comparisons will help elucidate the performance differences between the two device structures in terms of their transient behaviors. The fitted decay trace revealed the presence of two notable lifetime components: a longer rise time component (LRTC), $\tau_{1}$ , and a shorter rise time component (SRTC), $\tau_{2}$. After fitting the time response curve with a biexponential model, we found that the rise time and fall time for the InSe/1T$'$-WTe$_{2}$ configuration for  630 nm light are 42 ms and 126 ms, respectively, which is shown in figure-\ref{fig6}c. Similarly for InSe/Ti-Au configuration, we found that the rise time and fall time for the InSe/1T$'$-WTe$_{2}$ configuration for  630 nm light are 150 ms and 144 ms, respectively, which is shown in figure-\ref{fig6}d. For the UV configuration, the rise time and fall time for InSe/1T$'$-WTe$_{2}$ are 323 ms and 38 ms, respectively. In the NIR range, the rise time and fall time for InSe/1T$'$-WTe$_{2}$ are 96 ms and 671 ms. All these findings are summarized in Section S7 of the Supporting Information. In the UV range, for the InSe/Ti-Au configuration rise and fall times are 233 ms and 244 ms. For the NIR range, the rise time and fall time for InSe/Ti-Au are 384 ms and 131 ms, respectively, as shown in Section S3 of the Supporting Information. 
The responsivity of both devices reaches its peak in the DUV region (250 nm). To comprehend this phenomenon, it is essential to explore the underlying principles of photocurrent generation. This process is dictated by two fundamental mechanisms: the absorption of photons and the subsequent transport of charge carriers. As a result, the photocurrent ($I_{\text{ph}}$) is intrinsically linked to two pivotal physical parameters: quantum efficiency ($\eta_{q}$) and photoconductive gain ($\Gamma$). The latter, photoconductive gain ($\Gamma$), quantifies the efficiency of charge carrier dynamics and is determined by the ratio of carrier lifetime ($\tau_{\text{life}}$) to transit time ($\tau_{\text{transit}}$), which can be represented mathematically as
 \begin{equation}
\Gamma = \frac{V}{l^2} \tau_{life} \mu
\end{equation}
Here, $V$ represents the externally applied voltage, $l$ denotes the distance over which charge carriers travel, and $\mu$ corresponds to carrier mobility. Maximizing photoconductive gain can be accomplished by either shortening the transport path ($l$) or enhancing the applied voltage ($V$). Consequently, devices with minimized carrier transport distances exhibit superior photoconductive gain, leading to an increase in photocurrent generation. Despite the longer carrier transport distance of the InSe/1T$'$-WTe$_{2}$ device compared to the InSe/Ti-Au device, the former shows superior responsivity, highlighting the advantage of InSe/1T$'$-WTe$_{2}$. However, both $V$ and $l$ are external parameters. To assess the intrinsic photoconductive properties of InSe-based devices, it is crucial to examine fundamental characteristics such as carrier lifetime ($\tau_{\text{life}}$) and mobility ($\mu$), or more specifically, the $\tau_{\text{life}} \mu$ product, which is an intrinsic material parameter that governs carrier transport efficiency. To investigate the high photoconductive gain and responsivity in the DUV region for both InSe/1T$'$-WTe$_{2}$ and InSe/Ti-Au devices, the carrier lifetime ($\tau_{\text{life}}$) was analyzed using time-resolved current versus time graphs, as shown in figure-\ref{fig6}c and figure-\ref{fig6}d. The carrier charge trapping rise delay (CTRD) time was determined by fitting equation-\ref{equ 9} to the time-resolved current data for DUV, VIS, and NIR wavelengths for both devices. The results are shown in figures \ref{fig6}a, \ref{fig6}b, and figures S7c, S7d, S7g, and S7h of section S7 of Supporting Information. From the analysis, the carrier CTRD times for the InSe/1T$^{'}$-WTe$_{2}$ device in the DUV, VIS, and NIR regions were found to be 3.3 s, 1.4 s, and 1.2 s, respectively. Similarly, the carrier CTRD times for the InSe/Ti-Au device in these regions were 431 ms, 2.7 s, and 2.6 s, respectively. The remarkable photoconductive gain observed in the DUV region for both InSe-based devices stems from the interplay of enhanced carrier mobility and prolonged CTRD, or in other words, delayed carrier recombination . The elevated mobility arises from the inherently low effective electron mass ($m_e^* = 0.156 m_0$) characteristic of intrinsic n-type InSe, facilitating efficient charge transport. Meanwhile, the extended carrier CTRD time is predominantly governed by a surface-controlled photoconductivity (SCPC) mechanism, which becomes particularly significant in the DUV regime. This behavior aligns closely with the oxygen-sensitized photoconduction (OSPC) mechanism \cite{chen2012photoconduction, yang2018ultraefficient, soci2007zno}, further reinforcing the observed enhancement in photoconductive response. Figure-\ref{fig6}e shows the schematic diagram of different steps in OSPC. The oxygen-sensitized photoconduction (OSPC) model proposes that, at thermal equilibrium, oxygen molecules adhering to the material’s surface act as electron traps, leading to the following reaction:
\[
\mathrm{O_2(gas) + e^- \rightarrow O_2^-(adsorb)}
\]
This leads to the generation of negatively charged surface states and partial local surface defects, which give rise to surface band bending (SBB) (Step 1). When illuminated, electron-hole pairs (EHPs) are created (Step 2), and the separation of these carriers is governed by the SBB (Step 3). The inherent electric field at the surface directs the holes toward the surface, where they recombine with oxygen ions, releasing neutral oxygen molecules into the surrounding environment (\text{Step 4}). 
\[
[O_2^-(adsorb) + h^+ \rightarrow O_2(gas)]
\]
This process plays a pivotal role in shaping the material's optoelectronic properties and its interaction with surface states. As the majority of excess holes recombine at the surface, the remaining unpaired electrons exhibit prolonged lifetimes, thus dominating the photocurrent. The electron lifetime is primarily determined by the rate at which oxygen is adsorbed onto the surface (\text{Step 5}), which governs the overall process of electron retention and recombination. OSPC is a surface-dominant mechanism, meaning the region responsible for generating long-lifetime excess carriers is confined to the surface depletion region (SDR), particularly where surface band bending occurs. Since DUV light (250 nm) has a shorter penetration depth than VIS (630 nm) and NIR (950 nm) light, the carrier CTRD time for DUV in the InSe/1T$'$-WTe$_{2}$ device is 3.3 s, which is longer than the carrier CTRD times in the VIS and NIR regions.
\begin{table}[H]
\centering
\scalebox{0.75}{
\begin{tabular}{>{\raggedright\arraybackslash}p{5.2cm} 
                >{\raggedright\arraybackslash}p{4.5cm} 
                >{\centering\arraybackslash}p{4.2cm} 
                >{\centering\arraybackslash}p{3.2cm}}
\toprule
\textbf{Method \& Configuration} & \textbf{Photo Response Range (nm)} & \textbf{UV / Visible / NIR Trise/Tfall} & \textbf{Peak Responsivity} \\
\midrule

InSe/Au\cite{zhao2020role} & 530 nm & -- / 77 s / 3.2 s / -- & 0.09--0.11 AW\textsuperscript{-1} \\ \addlinespace

Few-layered InSe on SiO\textsubscript{2}/Si\cite{tamalampudi2014high} & 532 nm (peak), visible--NIR & -- / 50 ms / -- & 12 AW\textsuperscript{-1} \\ \addlinespace

CVD Graphene/InSe\cite{luo2015gate} (Graphene as contact) & 500 nm (peak), visible--NIR & -- / 128 µs / 220 µs / -- & 60 AW\textsuperscript{-1} \\ \addlinespace

PLD growth InSe/Metal\cite{yang2017wafer} & UV (peak), UV--NIR & 0.5 s / 1.7 s / -- & 27 AW\textsuperscript{-1} \\ \addlinespace

In/InSe/Au NP\cite{li2024high} & 637 nm & -- / 1.75 ms / -- & 15.2 AW\textsuperscript{-1} \\ \addlinespace

Gated InSe\cite{lei2014evolution} & 532 nm (peak), visible--NIR & -- / 488 µs / -- & 34.7 mA W\textsuperscript{-1} \\ \addlinespace

p-GaN/AlGaN/U-GaN/p-GaN/U-GaN HEMT\cite{wang2023light} & 365 nm (UV) & 248 µs / 584 µs / -- & 2.33 AW\textsuperscript{-1} \\ \addlinespace

MOCVD n$^+$-GaN/n$^-$-GaN/p-GaN/n$^+$-GaN MOSFET\cite{liu2023gan} & 365 nm (UV) & 1.34 ms / 4.72 ms / -- & 151.62 AW\textsuperscript{-1} \\ \addlinespace

p-GaN/u-GaN/AlN/AlGaN HEMT\cite{liu2024high} & 365 nm (UV) & 0.79 ms / 1.05 ms / -- & 532 AW\textsuperscript{-1} \\ \addlinespace

MOCVD-Grown β-Ga\textsubscript{2}O\textsubscript{3}\cite{khanzode2025highly} & 260 nm (DUV) & -- / -- / -- & 6.80 AW\textsuperscript{-1} \\ \addlinespace

ZnO-5\cite{sheoran2022high} & 253 nm (DUV) & 13 ms / -- / -- & 6.75 AW\textsuperscript{-1} \\ \addlinespace

\textbf{InSe/Ti-Au (This work)} & DUV (peak), UV--NIR & 233/244 ms (UV), 150/144 ms (Visible), 384/131 ms (NIR) & \textbf{3.64 AW\textsuperscript{-1}} \\ \addlinespace

\textbf{InSe/MBE grown 1T$'$-WTe\textsubscript{2} (This work)} & DUV (peak), DUV--NIR & 323/38 ms (UV), 42/126 ms (Visible), 96/671 ms (NIR) & \textbf{217.5 AW\textsuperscript{-1}} \\

\bottomrule
\end{tabular}
}
\caption{Comparison of various InSe and wide-bandgap semiconductor-based photodetector configurations showing their photo-response range, response times, and peak responsivities.}
\label{tab:photo_response_comparison}
\end{table}

As photocurrent is directly proportional to carrier lifetime, the highest responsivity is achieved in the DUV region. In the DUV region, a higher CTRD time is observed, indicating that carriers recombine later, meaning they spend a longer time before recombination. Figure-\ref{fig6}f and figure-\ref{fig6}g show the energy band diagrams of the InSe/1T$'$-WTe$_{2}$ device under drain-to-source bias when exposed to visible (VIS)/NIR light and DUV light, respectively. Due to the shorter penetration depth of DUV and the occurrence of OSPC, it generates a significant number of unpaired photogenerated electrons. This reduces the barrier height between the source and the channel more significantly compared to VIS/NIR light irradiation. Additionally, the depletion width becomes thinner due to the increased carrier concentration. As a combined effect, more carriers can travel from the source to the channel under DUV irradiation compared to VIS/NIR irradiation, as shown in figure-\ref{fig6}f and figure-\ref{fig6}g, respectively. Furthermore, OSPC results in the generation of more unpaired electrons in the DUV region, leading to a higher carrier lifetime of approximately 3.3 seconds compared to that under VIS and NIR light in the 1T$^\prime$-WTe\textsubscript{2}/InSe device. As a result, we observe a higher photoconductive gain or significant responsivity in the DUV region compared to VIS/NIR irradiation. In contrast, the InSe/Ti-Au device has a shorter carrier lifetime of 431 ms in the DUV region, which is lower than those for VIS and NIR light. Despite this, the responsivity in the DUV region remains high. DUV light generates a larger number of electron-hole pairs (EHPs) in the SDR, where photons are primarily absorbed near the surface, leading to more "effective holes" that interact with the charged surface states, resulting in higher "effective quantum efficiency." Conversely, VIS and NIR light, with greater penetration depths, generate fewer EHPs in the SDR, and the carrier lifetime in the neutral (bulk) region is significantly shorter due to direct recombination. Thus, we conclude that the contribution to the photocurrent from the neutral region is minimal, as the carriers in this region recombine rapidly. Stronger surface band bending (SBB) results in an extended electron lifetime, as unpaired electrons need to overcome the potential barrier before they can recombine or be captured by surface states, provided the oxygen capture rate is greater than the thermionic emission rate. In contrast, under conditions of higher light intensity, a reduced SBB leads to a shorter carrier lifetime due to more rapid recombination and a faster electron movement to the surface states. Now, comparing the responsivity of the InSe/1T$'$-WTe$_{2}$ device with the InSe/Ti-Au device, the InSe/1T$'$-WTe$_{2}$ device consistently outperforms the InSe/Ti-Au device. The first reason for this is the 2D/2D contact. This results in the InSe/1T$'$-WTe$_{2}$ device having a cleaner contact interface compared to the InSe/Ti-Au device, as evident from the fluctuations in the I–t curve of the InSe/Ti-Au device shown in figure-\ref{fig6}b. In contrast, the I–t curve of the 1T$^\prime$-WTe\textsubscript{2}/InSe device is very stable. The second reason is the significantly lower barrier height of the InSe/1T$'$-WTe$_{2}$ (0.393 eV, which is shown in figure-\ref{fig4}g) device compared to the InSe/Ti-Au device (0.750 eV, which is shown in figure-\ref{fig4}g). 
In Table \ref{tab:photo_response_comparison}, we summarize the photoresponse performance of InSe-based photodetectors from various literature sources, focusing on both metal and 2D contacts. From this table, we observe that our InSe/Ti-Au based photodetector exhibits a peak responsivity of 3.64 A/W in the UV range, which aligns well with findings reported in other studies. In this context, when we employ 1T$^\prime$-WTe\textsubscript{2} as a van der Waals electrode for InSe-based photodetectors, we notice a huge enhancement in responsivity, surpassing that of other InSe-based photodetectors. This configuration not only boosts responsivity but also accelerates the response time compared to our InSe/Ti-Au based photodetector. The superior performance of 1T$^\prime$-WTe\textsubscript{2} arises from its unique electronic properties and ability to form efficient contacts with the InSe material, paving the way for advanced optoelectronic applications. In addition to that, we also acknowledge that while the performance of our InSe/1T$^\prime$-WTe\textsubscript{2} device in the visible and near-infrared regions is modest, its responsivity and response speed in the ultraviolet (UV) range are significantly higher than those of many conventional two-terminal devices. Our device demonstrates markedly enhanced responsivity when compared to several traditional UV photodetectors, including those based on GaN, Ga \textsubscript{2}O\textsubscript{3}, and ZnO. Although a few reports have presented even higher responsivities and faster response times, such systems generally rely on complex architectures such as high-electron-mobility transistors (HEMTs) or metal–oxide–semiconductor field-effect transistors (MOSFETs), which inherently benefit from improved charge carrier separation and transport mechanisms. In contrast, our device adopts a simple two-terminal configuration, in which 1T$^\prime$-WTe\textsubscript{2} functions as a van der Waals (vdW) contact electrode on InSe. This work underscores how 2D electrodes can significantly enhance photodetector performance relative to conventional Ti/Au contacts, offering a more straightforward and scalable alternative to complex three-terminal or multilayered heterostructure-based designs.

\end{justify}

\section{Conclusion}
\begin{justify}
In summary, we synthesized highly crystalline, large-area 2D 1T$^\prime$-WTe\textsubscript{2} layers via molecular beam epitaxy (MBE) and demonstrated their use as electrical pads for broadband photodetection from deep ultraviolet (DUV) to near-infrared (NIR). A comparative analysis of InSe/1T$'$-WTe$_{2}$ and InSe/Ti-Au configurations revealed the impact of 1T$'$-WTe$_{2}$ on photodetector performance, driven by van der Waals interactions. The InSe/1T$'$-WTe$_{2}$ configuration exhibits superior photoresponsivity (0.143–217.58 A/W) and faster response times (42–323 ms rise, 38–671 ms fall) across the NIR–DUV spectrum, achieving its peak responsivity of 217.58 A/W in the DUV region. In contrast, the InSe/Ti-Au configuration shows lower responsivity ($8.65 \times 10^{-4}$–3.64 A/W) and slower response times. Enhanced UV responsivity in both devices is attributed to surface state effects, which induce band bending and modify electronic structure under UV exposure. This effect is more pronounced in InSe/1T$'$-WTe$_{2}$, where the already low energy barrier further decreases, enhancing carrier transport and photodetection efficiency. In addition, graphene exhibits a high dark current due to its zero band gap at the Dirac cone, leading to significant carrier leakage. To open a band gap, a complex, twisted structure of graphene is required. In contrast, 1T$^\prime$-WTe\textsubscript{2} possesses a small but finite band gap, which significantly reduces the dark current compared to graphene. Overall, the 1T$'$-WTe$_2$ van der Waals contact on InSe reduces the barrier height by half and lowers contact resistance by twenty-one times compared to metals, yielding sixty-fold higher responsivity and four-fold faster response. These results highlight 1T$'$-WTe$_2$ as an excellent contact material for 2D quantum electronics, enabling efficient electron injection and suppressed Fermi-level pinning. This property makes 1T$^\prime$-WTe\textsubscript{2} an excellent choice as a van der Waals electrode for advanced electronic devices, where low dark current is crucial for improved performance and energy efficiency.

\end{justify}

\section{\textbf{Experimental Section}}
\subsection{\textbf{\large{Materials Growth and Processing:}}}
\begin{justify}
1T$'$-WTe$_{2}$ thin films were grown using a Molecular Beam Epitaxy (MBE) system (Vinci Technologies, France) maintained at a base pressure of \( 5 \times 10^{-11} \, \text{mbar} \). High-purity tungsten (99.999\%) and tellurium (99.9999\%) sources were employed to ensure stoichiometric and high-quality film growth. Tungsten was evaporated via a Telemark e-beam evaporator, while tellurium was evaporated through an MBE KOMPONENTEN thermal cracker cell. A sapphire substrate was prepared by ultrasonic cleaning with acetone and isopropanol (IPA), followed by annealing at $700\,^{\circ}\mathrm{C}$ for 1.5 hours in an ultra-high vacuum (\( 5 \times 10^{-10} \, \text{mbar} \)) to remove surface contaminants. A Bayard-Alpert gauge was used to calibrate the flux of the materials, and a flux ratio of 1:80 (W:Te) was used for the 1T$^\prime$-WTe\textsubscript{2} growth. The deposition temperature was optimized at $450\,^{\circ}\mathrm{C}$, with a controlled growth rate of approximately 90 minutes per monolayer to achieve smooth surfaces and good crystalline 1T$^\prime$-WTe\textsubscript{2} film. Real-time Reflection High-Energy Electron Diffraction (RHEED) monitoring was conducted at 12 kV and 1.48 A. \vspace{12pt}
\end{justify}

\subsection{\textbf{\large{Etching and device fabrication details}}}
\begin{justify}
The etching of 1T$^\prime$-WTe\textsubscript{2} was carried out using an HHV (Sara) reactive ion etching (RIE) system. After patterning 1T$^\prime$-WTe\textsubscript{2} with a laser writer (Microtech, LW405), the sample was baked at $80\,^{\circ}\mathrm{C}$ for 25 mins. The sample was then placed inside the RIE, where we waited for 2 hours and 15 minutes to achieve a vacuum of \( 1.5 \times 10^{-6} \, \text{mbar} \). Once this pressure was reached, an optimized etching process was performed using a gas mixture of SF$_{6}$, O$_{2}$, and argon, in a ratio of 45:12:5 sccm, respectively. The etching process was performed at 50 watts of power for 6 minutes at heating temperature $90\,^{\circ}\mathrm{C}$. Following the etching, we obtained a well defined 1T$^\prime$-WTe\textsubscript{2} electrical pad, ready for the transfer of InSe. To transfer InSe, we used a Holmarc 2D transfer system (Model No: HO-2DTS-DM-02). First, we transferred the InSe crystal onto PDMS (Gelpack PF film), attached it to a glass slide, and mounted it on the stamping stage of the 2D transfer system. We then waited for the sample stage to reach a temperature of $85\,^{\circ}\mathrm{C}$. Once the target temperature was reached, we carefully lowered the stamping glass slide onto the substrate to transfer the micron-sized InSe crystal into the micron-sized gap of 1T$^\prime$-WTe\textsubscript{2}. \vspace{8pt}
\end{justify}

\subsection{\textbf{\large{Characterization Techniques:}}}
\subsubsection{\textbf{\large{Surface Topography and KPFM: }}}
\begin{justify}
Atomic Force Microscopy (AFM) was performed using a Bruker Dimension Icon XR system. For general imaging, an RTESPA-300 probe (8 nm tip radius) was used, while an SCM-PIT-V2 probe (25 nm tip radius) was applied in Kelvin Probe Force Microscopy (KPFM) mode.

\end{justify}

 \subsubsection{\textbf{\large{Phase and Composition:}}}
 \begin{justify}
 Raman spectroscopy (Renishaw inVia) with 532 nm laser excitation and 0.5 cm-1 spectral resolution provided phase analysis. Chemical composition and valence band spectra were assessed via X-ray Photoelectron Spectroscopy (XPS) using an AXIS Supra system (Kratos Analytical) with an Al K-alpha source (1486.7 eV) and a resolution of 0.55 eV.
\end{justify}

 \subsection{\textbf{\large{Optical and Electrical Properties:}}}
\begin{justify}
 UV-Vis-NIR spectroscopy was conducted on a Lambda 1050 system (PerkinElmer) equipped with a triple detector module, yielding a spectral resolution of 0.05 nm (UV-Vis) and 0.20 nm (NIR) with 1 nm step intervals. Electrical properties, including current-voltage (I-V) measurements, were evaluated using a DC probe station (EverBeing EB6) integrated with a Keithley Semiconductor Characterization System (SCS-4200). The speed of both our devices was measured by fitting an exponential curve to the I-t curve, which we manually obtained using this semiconductor characterization system. However, our existing system has a limitation on data points, so the actual speed of both devices might be higher than the reported value in this study. Nevertheless, here we aim to convey that the 1T$^\prime$-WTe\textsubscript{2} electrode-based device outperforms conventional metal-based devices in all aspects of performance. In addition, for the TLM measurements, we used a Keithley 2450 source-measure unit with automation implemented via Python. First, we started the sweep in the nanoampere range and recorded the device response, gradually increasing the current while setting a safe compliance limit to protect the device. Finally, in the microampere range, we observed stable outputs, and all measurements were carried out after several trials and careful monitoring of the device response. All measurements were performed with the Number of Power Line Cycles (NPLC) set to 10, in order to reduce measurement noise and improve accuracy.

\end{justify}

\begin{justify}
\end{justify}

\medskip
\textbf{Supporting Information}
\begin{justify}
\textbf{S1:} EDX of molecular Beam epitaxy Growth of 1T$^\prime$-WTe\textsubscript{2}. \textbf{S2:} Dry etching of WTe$_{2}$ to serve as an electrical pad. \textbf{S3:} Dry transfer of InSe on patterned WTe$_2$ contact. \textbf{S4:} Extraction of experimental barrier height. \textbf{S5:} Heterostructure band diagram. \textbf{S6:} Experimental extraction of contact resistance.\textbf{S7:} InSe/Ti-Au and 1T$^\prime$-WTe\textsubscript{2}/InSe based device and characteristics \textbf{S8:} Time response of both InSe/1T$^\prime$-WTe\textsubscript{2} and InSe/Ti-Au devices.\vspace{2pt} 
\end{justify}

\medskip
\textbf{Acknowledgements}
\begin{justify}
The authors express their gratitude to the Defence Research and Development Organisation (DRDO) and the Department of Science and Technology (DST) for their financial support. They also extend their thanks to the Ministry of Human Resource Development (MHRD), India, for funding this work through the 'Grand Challenge Project on MBE Growth of 2D Materials' (MI01800G) and MoE-STARS/STARS-2/2023-061. Additionally, the authors acknowledge the Centre Research Facility (CRF), Nano Research Facility (NRF), and Centre for Applied Research in Electronics (CARE) at IIT Delhi for their assistance in providing various characterization and fabrication facilities essential to this research. BK acknowledges the financial assistance from the Prime Minister Research Fellows (PMRF) Scheme (PMRF ID: 1401633), India.\vspace{8pt}. 
\end{justify}
\par 

\medskip
\textbf{Conflict of Interest}
\begin{justify}
The Authors declare no conflict of interest.
\end{justify}

\medskip
\textbf{Author Contributions}
\begin{justify}
B.K and S.K contributed equally to this paper.
\end{justify}

\medskip
\textbf{Data Availability}
\begin{justify}
The data are available from the corresponding author upon reasonable request.
\end{justify}

\textbf{Correspondence} and inquiries for materials should be directed to S.D.

\clearpage

\bibliographystyle{MSP}
\bibliography{BK_Main}

\begin{thebibliography}{10}
\providecommand{\url}[1]{\texttt{#1}}
\providecommand{\urlprefix}{URL }

\bibitem{liu2021transferred}
L.~Liu, L.~Kong, Q.~Li, C.~He, L.~Ren, Q.~Tao, X.~Yang, J.~Lin, B.~Zhao, Z.~Li, et~al.,
\newblock \emph{Nature Electronics} \textbf{2021}, \emph{4}, 5 342.

\bibitem{cusati2017electrical}
T.~Cusati, G.~Fiori, A.~Gahoi, V.~Passi, M.~C. Lemme, A.~Fortunelli, G.~Iannaccone,
\newblock \emph{Scientific reports} \textbf{2017}, \emph{7}, 1 5109.

\bibitem{mootheri2020graphene}
V.~Mootheri, G.~Arutchelvan, S.~Banerjee, S.~Sutar, A.~Leonhardt, M.-E. Boulon, C.~Huyghebaert, M.~Houssa, I.~Asselberghs, I.~Radu, et~al.,
\newblock \emph{2D Materials} \textbf{2020}, \emph{8}, 1 015003.

\bibitem{chen2015high}
Z.~Chen, J.~Biscaras, A.~Shukla,
\newblock \emph{Nanoscale} \textbf{2015}, \emph{7}, 14 5981.

\bibitem{xia2009ultrafast}
F.~Xia, T.~Mueller, Y.-m. Lin, A.~Valdes-Garcia, P.~Avouris,
\newblock \emph{Nature nanotechnology} \textbf{2009}, \emph{4}, 12 839.

\bibitem{wang2012electronics}
Q.~H. Wang, K.~Kalantar-Zadeh, A.~Kis, J.~N. Coleman, M.~S. Strano,
\newblock \emph{Nature nanotechnology} \textbf{2012}, \emph{7}, 11 699.

\bibitem{bonaccorso2010graphene}
F.~Bonaccorso, Z.~Sun, T.~Hasan, A.~C. Ferrari,
\newblock \emph{Nature photonics} \textbf{2010}, \emph{4}, 9 611.

\bibitem{nematollahi2019weyl}
F.~Nematollahi, S.~A. Oliaei~Motlagh, V.~Apalkov, M.~I. Stockman,
\newblock \emph{Physical Review B} \textbf{2019}, \emph{99}, 24 245409.

\bibitem{wang2022hybrid}
L.~Wang, L.~Han, W.~Guo, L.~Zhang, C.~Yao, Z.~Chen, Y.~Chen, C.~Guo, K.~Zhang, C.-N. Kuo, et~al.,
\newblock \emph{Light: Science \& Applications} \textbf{2022}, \emph{11}, 1 53.

\bibitem{ma2019nonlinear}
J.~Ma, Q.~Gu, Y.~Liu, J.~Lai, P.~Yu, X.~Zhuo, Z.~Liu, J.-H. Chen, J.~Feng, D.~Sun,
\newblock \emph{Nature materials} \textbf{2019}, \emph{18}, 5 476.

\bibitem{yang2023centimeter}
Q.~Yang, X.~Wang, Z.~He, Y.~Chen, S.~Li, H.~Chen, S.~Wu,
\newblock \emph{Advanced Science} \textbf{2023}, \emph{10}, 17 2205609.

\bibitem{chan2017photocurrents}
C.-K. Chan, N.~H. Lindner, G.~Refael, P.~A. Lee,
\newblock \emph{Physical Review B} \textbf{2017}, \emph{95}, 4 041104.

\bibitem{yin2012single}
Z.~Yin, H.~Li, H.~Li, L.~Jiang, Y.~Shi, Y.~Sun, G.~Lu, Q.~Zhang, X.~Chen, H.~Zhang,
\newblock \emph{ACS nano} \textbf{2012}, \emph{6}, 1 74.

\bibitem{butler2013progress}
S.~Z. Butler, S.~M. Hollen, L.~Cao, Y.~Cui, J.~A. Gupta, H.~R. Guti{\'e}rrez, T.~F. Heinz, S.~S. Hong, J.~Huang, A.~F. Ismach, et~al.,
\newblock \emph{ACS nano} \textbf{2013}, \emph{7}, 4 2898.

\bibitem{lei2015atomically}
S.~Lei, F.~Wen, L.~Ge, S.~Najmaei, A.~George, Y.~Gong, W.~Gao, Z.~Jin, B.~Li, J.~Lou, et~al.,
\newblock \emph{Nano letters} \textbf{2015}, \emph{15}, 5 3048.

\bibitem{feng2015ultrahigh}
W.~Feng, J.-B. Wu, X.~Li, W.~Zheng, X.~Zhou, K.~Xiao, W.~Cao, B.~Yang, J.-C. Idrobo, L.~Basile, et~al.,
\newblock \emph{Journal of Materials Chemistry C} \textbf{2015}, \emph{3}, 27 7022.

\bibitem{zhao2020role}
Q.~Zhao, W.~Wang, F.~Carrascoso-Plana, W.~Jie, T.~Wang, A.~Castellanos-Gomez, R.~Frisenda,
\newblock \emph{Materials horizons} \textbf{2020}, \emph{7}, 1 252.

\bibitem{tamalampudi2014high}
S.~R. Tamalampudi, Y.-Y. Lu, R.~K. U, R.~Sankar, C.-D. Liao, C.-H. Cheng, F.~C. Chou, Y.-T. Chen,
\newblock \emph{Nano letters} \textbf{2014}, \emph{14}, 5 2800.

\bibitem{luo2015gate}
W.~Luo, Y.~Cao, P.~Hu, K.~Cai, Q.~Feng, F.~Yan, T.~Yan, X.~Zhang, K.~Wang,
\newblock \emph{Advanced Optical Materials} \textbf{2015}, \emph{3}, 10 1418.

\bibitem{lei2021ambipolar}
T.~Lei, H.~Tu, W.~Lv, H.~Ma, J.~Wang, R.~Hu, Q.~Wang, L.~Zhang, B.~Fang, Z.~Liu, et~al.,
\newblock \emph{ACS Applied Materials \& Interfaces} \textbf{2021}, \emph{13}, 42 50213.

\bibitem{zhao2018highly}
S.~Zhao, J.~Wu, K.~Jin, H.~Ding, T.~Li, C.~Wu, N.~Pan, X.~Wang,
\newblock \emph{Advanced Functional Materials} \textbf{2018}, \emph{28}, 34 1802011.

\bibitem{patil2022self}
C.~Patil, C.~Dong, H.~Wang, B.~M. Nouri, S.~Krylyuk, H.~Zhang, A.~V. Davydov, H.~Dalir, V.~J. Sorger,
\newblock \emph{Photonics Research} \textbf{2022}, \emph{10}, 7 A97.

\bibitem{shang2020mixed}
H.~Shang, H.~Chen, M.~Dai, Y.~Hu, F.~Gao, H.~Yang, B.~Xu, S.~Zhang, B.~Tan, X.~Zhang, et~al.,
\newblock \emph{Nanoscale Horizons} \textbf{2020}, \emph{5}, 3 564.

\bibitem{ma2022ultrasensitive}
H.~Ma, Y.~Xing, J.~Han, B.~Cui, T.~Lei, H.~Tu, B.~Guan, Z.~Zeng, B.~Zhang, W.~Lv,
\newblock \emph{Advanced Optical Materials} \textbf{2022}, \emph{10}, 5 2101772.

\bibitem{sun2021anti}
Y.~Sun, W.~Gao, X.~Li, C.~Xia, H.~Chen, L.~Zhang, D.~Luo, W.~Fan, N.~Huo, J.~Li,
\newblock \emph{Journal of Materials Chemistry C} \textbf{2021}, \emph{9}, 32 10372.

\bibitem{xiong2021high}
J.~Xiong, Y.~Sun, L.~Wu, W.~Wang, W.~Gao, N.~Huo, J.~Li,
\newblock \emph{Advanced Optical Materials} \textbf{2021}, \emph{9}, 20 2101017.

\bibitem{gao2022low}
P.~Gao, M.~Yang, C.~Wang, H.~Li, B.~Yang, Z.~Zheng, N.~Huo, W.~Gao, D.~Luo, J.~Li,
\newblock \emph{Nanoscale} \textbf{2022}, \emph{14}, 39 14603.

\bibitem{zhang2023type}
Z.~Zhang, L.~Han, Z.~Dan, H.~Li, M.~Yang, Y.~Sun, Z.~Zheng, N.~Huo, D.~Luo, W.~Gao, et~al.,
\newblock \emph{ACS Applied Nano Materials} \textbf{2023}, \emph{6}, 6 4573.

\bibitem{dan2021improved}
Z.~Dan, C.~Wang, W.~Gao, K.~Shu, L.~Wu, W.~Wang, Q.~Zhao, X.~Liu, X.~Liu, N.~Huo, et~al.,
\newblock \emph{APL Materials} \textbf{2021}, \emph{9}, 8.

\bibitem{liu2022selectively}
J.~Liu, Q.~Hao, H.~Gan, P.~Li, B.~Li, Y.~Tu, J.~Zhu, D.~Qi, Y.~Chai, W.~Zhang, et~al.,
\newblock \emph{Laser \& Photonics Reviews} \textbf{2022}, \emph{16}, 11 2200338.

\bibitem{shang2022carrier}
H.~Shang, Y.~Hu, F.~Gao, M.~Dai, S.~Zhang, S.~Wang, D.~Ouyang, X.~Li, X.~Song, B.~Gao, et~al.,
\newblock \emph{ACS nano} \textbf{2022}, \emph{16}, 12 21293.

\bibitem{ma2024vertical}
J.~Ma, J.~Wang, Q.~Chen, S.~Chen, M.~Yang, Y.~Sun, Z.~Zheng, N.~Huo, Y.~Yan, J.~Li, et~al.,
\newblock \emph{Advanced Electronic Materials} \textbf{2024}, \emph{10}, 1 2300672.

\bibitem{xie2021gate}
Y.~Xie, E.~Wu, G.~Geng, D.~Zhang, X.~Hu, J.~Liu,
\newblock \emph{Applied Physics Letters} \textbf{2021}, \emph{118}, 13.

\bibitem{wang2022weyl}
J.~Wang, H.~Wang, Q.~Chen, L.~Qi, Z.~Zheng, N.~Huo, W.~Gao, X.~Wang, J.~Li,
\newblock \emph{Applied Physics Letters} \textbf{2022}, \emph{121}, 10.

\bibitem{li2021tunable}
L.~Li, G.~Zhang, H.~Wu, L.~Yang, P.~Gao, S.~Zhang, X.~Wen, W.~Zhang, H.~Chang \textbf{2021}.

\bibitem{naylor2017large}
C.~H. Naylor, W.~M. Parkin, Z.~Gao, H.~Kang, M.~Noyan, R.~B. Wexler, L.~Z. Tan, Y.~Kim, C.~E. Kehayias, F.~Streller, et~al.,
\newblock \emph{2D Materials} \textbf{2017}, \emph{4}, 2 021008.

\bibitem{zhou2017direct}
Y.~Zhou, H.~Jang, J.~M. Woods, Y.~Xie, P.~Kumaravadivel, G.~A. Pan, J.~Liu, Y.~Liu, D.~G. Cahill, J.~J. Cha,
\newblock \emph{Advanced Functional Materials} \textbf{2017}, \emph{27}, 8 1605928.

\bibitem{lin2024dramatically}
W.-H. Lin, C.-S. Li, C.-I. Wu, G.~R. Rossman, H.~A. Atwater, N.-C. Yeh,
\newblock \emph{Advanced Science} \textbf{2024}, \emph{11}, 2 2304890.

\bibitem{li2020molecular}
H.~Li, A.~Chen, L.~Wang, W.~Ren, S.~Lu, B.~Yang, Y.-P. Jiang, F.-S. Li,
\newblock \emph{Applied Physics Letters} \textbf{2020}, \emph{117}, 16.

\bibitem{walsh2017w}
L.~A. Walsh, R.~Yue, Q.~Wang, A.~T. Barton, R.~Addou, C.~M. Smyth, H.~Zhu, J.~Kim, L.~Colombo, M.~J. Kim, et~al.,
\newblock \emph{2D Materials} \textbf{2017}, \emph{4}, 2 025044.

\bibitem{khan2024observation}
B.~Khan, A.~Mukherjee, Y.~M. Georgiev, J.-P. Colinge, S.~Ghosh, S.~Das,
\newblock \emph{arXiv preprint arXiv:2403.07324} \textbf{2024}.

\bibitem{buchkov2021anisotropic}
K.~Buchkov, R.~Todorov, P.~Terziyska, M.~Gospodinov, V.~Strijkova, D.~Dimitrov, V.~Marinova,
\newblock \emph{Nanomaterials} \textbf{2021}, \emph{11}, 9 2262.

\bibitem{kim2016anomalous}
Y.~Kim, Y.~I. Jhon, J.~Park, J.~H. Kim, S.~Lee, Y.~M. Jhon,
\newblock \emph{Nanoscale} \textbf{2016}, \emph{8}, 4 2309.

\bibitem{xiao2024photoelectric}
Y.~Xiao, K.~Luo, Q.~Kao, Y.~Fu, W.~Jiang, L.~Cao,
\newblock \emph{Surfaces and Interfaces} \textbf{2024}, \emph{44} 103670.

\bibitem{brown1966crystal}
B.~E. Brown,
\newblock \emph{Acta Crystallographica} \textbf{1966}, \emph{20}, 2 268.

\bibitem{augustin2000electronic}
J.~Augustin, V.~Eyert, T.~B{\"o}ker, W.~Frentrup, H.~Dwelk, C.~Janowitz, R.~Manzke,
\newblock \emph{Physical Review B} \textbf{2000}, \emph{62}, 16 10812.

\bibitem{lee2015tungsten}
C.-H. Lee, E.~C. Silva, L.~Calderin, M.~A.~T. Nguyen, M.~J. Hollander, B.~Bersch, T.~E. Mallouk, J.~A. Robinson,
\newblock \emph{Scientific reports} \textbf{2015}, \emph{5}, 1 10013.

\bibitem{zheng2016quantum}
F.~Zheng, C.~Cai, S.~Ge, X.~Zhang, X.~Liu, H.~Lu, Y.~Zhang, J.~Qiu, T.~Taniguchi, K.~Watanabe, et~al.,
\newblock \emph{arXiv preprint arXiv:1605.04656} \textbf{2016}.

\bibitem{weng2023polarization}
X.~Weng, L.~Qi, W.~Tang, M.~A. Iqbal, C.~Kang, K.~Wu, Y.-J. Zeng,
\newblock \emph{RSC advances} \textbf{2023}, \emph{13}, 48 33588.

\bibitem{li2022interfacial}
J.~Li, L.~Wang, Y.~Chen, Y.~Li, H.~Zhu, L.~Li, L.~Tong,
\newblock \emph{Nanomaterials} \textbf{2022}, \emph{13}, 1 147.

\bibitem{mleczko2016high}
M.~J. Mleczko, R.~L. Xu, K.~Okabe, H.-H. Kuo, I.~R. Fisher, H.-S.~P. Wong, Y.~Nishi, E.~Pop,
\newblock \emph{ACS nano} \textbf{2016}, \emph{10}, 8 7507.

\bibitem{chen2017simple}
K.~Chen, Z.~Chen, X.~Wan, Z.~Zheng, F.~Xie, W.~Chen, X.~Gui, H.~Chen, W.~Xie, J.~Xu,
\newblock \emph{Advanced Materials} \textbf{2017}, \emph{29}, 38 1700704.

\bibitem{yang2017wafer}
Z.~Yang, W.~Jie, C.-H. Mak, S.~Lin, H.~Lin, X.~Yang, F.~Yan, S.~P. Lau, J.~Hao,
\newblock \emph{ACS nano} \textbf{2017}, \emph{11}, 4 4225.

\bibitem{maiti2020strain}
R.~Maiti, C.~Patil, M.~Saadi, T.~Xie, J.~Azadani, B.~Uluutku, R.~Amin, A.~Briggs, M.~Miscuglio, D.~Van~Thourhout, et~al.,
\newblock \emph{Nature Photonics} \textbf{2020}, \emph{14}, 9 578.

\bibitem{melitz2011kelvin}
W.~Melitz, J.~Shen, A.~C. Kummel, S.~Lee,
\newblock \emph{Surface science reports} \textbf{2011}, \emph{66}, 1 1.

\bibitem{kim2017fermi}
C.~Kim, I.~Moon, D.~Lee, M.~S. Choi, F.~Ahmed, S.~Nam, Y.~Cho, H.-J. Shin, S.~Park, W.~J. Yoo,
\newblock \emph{ACS nano} \textbf{2017}, \emph{11}, 2 1588.

\bibitem{jeon2018epitaxial}
J.~Jeon, Y.~Park, S.~Choi, J.~Lee, S.~S. Lim, B.~H. Lee, Y.~J. Song, J.~H. Cho, Y.~H. Jang, S.~Lee,
\newblock \emph{ACS nano} \textbf{2018}, \emph{12}, 1 338.

\bibitem{schroder2015semiconductor}
D.~K. Schroder,
\newblock \emph{Semiconductor material and device characterization},
\newblock John Wiley \& Sons, \textbf{2015}.

\bibitem{khan2024unveiling}
T.~Khan, K.~Arora, R.~Agarwal, P.~K. Muduli, Y.-H. Chu, R.~H. Horng, R.~Singh,
\newblock \emph{ACS Applied Materials \& Interfaces} \textbf{2024}.

\bibitem{kaushik20212d}
S.~Kaushik, R.~Singh,
\newblock \emph{Advanced Optical Materials} \textbf{2021}, \emph{9}, 11 2002214.

\bibitem{kandar2025scalable}
S.~Kandar, K.~Bhatt, S.~Tiwari, N.~Chaudhary, T.~Khan, A.~Kapoor, R.~Singh,
\newblock \emph{Journal of Materials Chemistry C} \textbf{2025}.

\bibitem{chen2012photoconduction}
R.~Chen, C.~Chen, H.~Tsai, W.~Wang, Y.~Huang,
\newblock \emph{The Journal of Physical Chemistry C} \textbf{2012}, \emph{116}, 6 4267.

\bibitem{yang2018ultraefficient}
H.-W. Yang, H.-F. Hsieh, R.-S. Chen, C.-H. Ho, K.-Y. Lee, L.-C. Chao,
\newblock \emph{ACS applied materials \& interfaces} \textbf{2018}, \emph{10}, 6 5740.

\bibitem{soci2007zno}
C.~Soci, A.~Zhang, B.~Xiang, S.~A. Dayeh, D.~Aplin, J.~Park, X.~Bao, Y.-H. Lo, D.~Wang,
\newblock \emph{Nano letters} \textbf{2007}, \emph{7}, 4 1003.

\bibitem{li2024high}
L.~Li, Q.~Wu, C.~Wang, Z.~Cai, L.~Lin, X.~Gu, K.~K. Ostrikov, H.~Nan, S.~Xiao,
\newblock \emph{ACS Photonics} \textbf{2024}.

\bibitem{lei2014evolution}
S.~Lei, L.~Ge, S.~Najmaei, A.~George, R.~Kappera, J.~Lou, M.~Chhowalla, H.~Yamaguchi, G.~Gupta, R.~Vajtai, et~al.,
\newblock \emph{ACS nano} \textbf{2014}, \emph{8}, 2 1263.

\bibitem{wang2023light}
Y.~Wang, C.~Liu, H.~Qian, H.~Liu, L.~Han, X.~Wang, W.~Gao, J.~Li,
\newblock \emph{Optics Letters} \textbf{2023}, \emph{48}, 16 4376.

\bibitem{liu2023gan}
H.~Liu, Y.~Wang, C.~Liu, W.~Gao, L.~Han, X.~Wang, J.~Li,
\newblock \emph{ACS Applied Optical Materials} \textbf{2023}, \emph{1}, 8 1485.

\bibitem{liu2024high}
C.~Liu, Y.~Wang, H.~Liu, H.~Qian, L.~Han, X.~Wang, Z.~Zheng, W.~Gao, J.~Li,
\newblock \emph{ACS Applied Electronic Materials} \textbf{2024}, \emph{6}, 2 1347.

\bibitem{khanzode2025highly}
P.~M. Khanzode, S.~J. Shaikh, D.~I. Halge, M.~Y. Thabit, A.~B. Rahman, V.~N. Narwade, S.~S. Dahiwale, K.~A. Bogle,
\newblock \emph{ACS Applied Engineering Materials} \textbf{2025}.

\bibitem{sheoran2022high}
H.~Sheoran, S.~Fang, F.~Liang, Z.~Huang, S.~Kaushik, N.~Manikanthababu, X.~Zhao, H.~Sun, R.~Singh, S.~Long,
\newblock \emph{ACS Applied Materials \& Interfaces} \textbf{2022}, \emph{14}, 46 52096.

\end{thebibliography}
\end{document}